\documentclass[aps, pra, 10pt]{revtex4-2} 

\usepackage{graphicx}
\usepackage{amsmath}
\usepackage{indentfirst}
\usepackage{here}
\usepackage{bm}

\begin{document}

\title{Quantum electrometry in a silicon carbide power device}

\author{Yuichi Yamazaki\textsuperscript{1}}
\author{Akira Kiyoi\textsuperscript{2}}
\author{Naoyuki Kawabata\textsuperscript{2}}
\author{Yuki Watanabe\textsuperscript{2}}
\author{Ryosuke Akashi\textsuperscript{1}}
\author{Shunsuke Daimon\textsuperscript{1}}
\author{Nobumasa Miyawaki\textsuperscript{1}}
\author{Yu-ichiro Matsushita\textsuperscript{1,3,4}}
\author{Makoto Kohda\textsuperscript{1,5,6}}
\author{Takeshi Ohshima\textsuperscript{1,5}}

 \affiliation{$^1$National Institutes for Quantum Science and Technology (QST), Takasaki, Gunma 370-1292, Japan}
 \affiliation{$^2$Advanced Technology R\&D Center, Mitsubishi Electric Corporation, Amagasaki, Hyogo 661-8661, Japan}
 \affiliation{$^3$Department of Physics, The University of Tokyo, Hongo, Bunkyo-ku, Tokyo, Japan}
\affiliation{$^4$Quemix Inc., Taiyo Life Nihombashi Building, 2-11-2, Nihombashi Chuo-ku, Tokyo, Japan}
\affiliation{$^5$Department of Materials Science, Tohoku University, Sendai, Miyagi 980-8579, Japan}
\affiliation{$^6$Department of Materials Science and Engineering, University of Washington, Seattle, Washington 98195, USA}

\email{yamazaki.yuichi@qst.go.jp}

\begin{abstract}
For high-bias operation devices such as silicon carbide (SiC) power devices, early detection of failure mechanisms is essential to ensure reliability. This requires a method to map high electric fields with high spatial resolution, which has not been realized until now. Here we report that the silicon vacancy (V\textsubscript{Si}) in SiC has outstanding characteristics for detecting electric fields applied in various directions within a high-biased SiC device. V\textsubscript{Si} exhibits an equivalent response to electric field components parallel (\emph{E}\textsubscript{\textbar\textbar{}}) and perpendicular
(\emph{E}\textsubscript{$\bot$}) to the \emph{c}-axis, a feature unique among quantum sensors, and the responsiveness to
\emph{E}\textsubscript{\textbar\textbar{}} and \emph{E}\textsubscript{$\bot$} enables detection of arbitrary electric fields encountered in cutting-edge SiC power devices. We confirmed high electric field detection of $\sim$2.3 MV/cm, which is $\sim$90\% of the breakdown electric field of a 4H-SiC with typical carrier concentration. Selectively formed V\textsubscript{Si} enables high-resolution mapping of electric field distribution. V\textsubscript{Si}-based quantum sensors bring data-driven research and development methodologies as well as device degradation diagnosis.
\end{abstract}

\maketitle

\section*{Introduction}
The rapid adoption of artificial-intelligence technologies and electric-vehicle platforms has resulted in an unprecedented rise in electrical power consumption, thereby elevating power device performance to a critical bottleneck in modern energy-conversion systems. Silicon carbide (SiC) has emerged as a leading material for advanced power electronics offering higher efficiency, greater thermal conductivity, and higher voltage operation \cite{a1}.

The reliability of such high-biased devices depends on the early detection of failure mechanisms such as hot-spot formation, premature breakdown, and trap-induced leakage, all of which manifest as localized distortions of the internal electric field distribution \cite{a2, a3}. Therefore, high-resolution mapping of the electric field inside a working device would provide a powerful diagnostic tool and a pathway to improve device design. Conventional probing techniques, such as Kelvin probe force microscopy \cite{a4}, differential phase contrast scanning transmission electron microscopy \cite{a5, a6} and nonlinear optical approaches\cite{a7, a8}, suffer from issues such as macroscopic spatial averaging, invasive contact requirements, and limitations in sensitivity to nanoscale electric field gradients preceding catastrophic failure.

Quantum sensors based on point-defect spins in wide-bandgap
semiconductors (spin defects) \cite{a9, a10, a11, a12, a13, a14}, which can measure various
physical quantities such as a magnetic field, an electric field, and
temperature with high sensitivity and high spatial resolution
\cite{a15, a16, a17, a18, a19, a20, a21}, have recently emerged as a promising solution to these
limitations. Optically addressable spin defects such as divacancy
(V\textsubscript{Si}V\textsubscript{C}) and silicon vacancy
(V\textsubscript{Si}) in SiC exhibit electric-field-dependent shifts of
their spin resonance frequencies via the Stark effect \cite{a22, a23}. By
performing optically detected magnetic resonance (ODMR) measurements on
these defects embedded into a device, a local electric field can be
inferred with nanometre-scale spatial resolution and high sensitivity,
opening the prospect of non-invasive, three-dimensional field imaging
inside power devices in operando.

A demonstration of electric field sensing using
V\textsubscript{Si}V\textsubscript{C} has already been reported
\cite{a22}. However, sensitivity of the
V\textsubscript{Si}V\textsubscript{C} defect to electric fields is
intrinsically anisotropic: the intrinsic axis that connects the two
vacancies (the V\textsubscript{Si}-V\textsubscript{C} axis) imposes a
strongly anisotropic potential on the electrons (Fig.\,1a), and the
Stark tensor is anisotropic, with distinct components parallel and
perpendicular to the V\textsubscript{Si}-V\textsubscript{C} axis. As a
result, the sensitivity to electric field components parallel and
perpendicular to the V\textsubscript{Si}-V\textsubscript{C} axis differs
significantly \cite{a22}. Moreover, the
V\textsubscript{Si}V\textsubscript{C} requires high temperature
annealing (typically \textgreater{} 800 $^\circ$C) for defect formation and
reliable ODMR measurement is limited to cryogenic temperatures, which
precludes straightforward integration with power devices.

In contrast, V\textsubscript{Si} possesses an intrinsically more
isotropic electrostatic environment. The vacancy, though under weak
uniaxial field by the 4H stacking, is surrounded by four carbon atoms
arranged in the tetrahedral coordination as shown in Fig.\,1b.
Consequently, V\textsubscript{Si} is expected to exhibit high
sensitivity to electric field components in all spatial directions,
enabling three-dimensional electric field detection. Combining this
isotropy with the fact that V\textsubscript{Si} exhibits robust ODMR
from room temperature to $\sim$590\,K \cite{a24} makes it ideally
suited for in-situ diagnostics of SiC power devices. A recent
proof-of-concept study has proven that a V\textsubscript{Si}-based
quantum sensor can detect electric field in a working device \cite{a23}.
Yet, the direction of the electric field was restricted to a single
axis, which is parallel to the quantization axis of V\textsubscript{Si}
(\emph{E}\textsubscript{\textbar\textbar{}}), and the measured electric
field was more than an order of magnitude smaller than the typical
breakdown electric field of SiC (=\,2-3\,MV/cm). To analyze failures on
devices in practical use, the measurement technique must be able to
operate under high electric fields of approximately 2 MV/cm, which are
typical under normal operating conditions. Furthermore, to measure
cutting-edge devices like a trench gate MOSFET and super-junction device
\cite{a25,a26, a27}, which generate electric field components perpendicular to
the \emph{c}-axis (\emph{E}\textsubscript{$\bot$}), methods capable of
detecting not only \emph{E}\textsubscript{\textbar\textbar{}} but also
\emph{E}\textsubscript{$\bot$} are essential.

In this article, we demonstrate that V\textsubscript{Si}-based quantum
sensors provide a unique capability for detecting electric fields of
arbitrary orientation in a high-biased SiC power device. By embedding
V\textsubscript{Si} ensemble at selected positions (Fig.\,1c), we first
quantify the electric dipole moments parallel
(\emph{d}\textsubscript{\textbar\textbar{}}) and perpendicular
(\emph{d}\textsubscript{$\bot$}) to the crystal \emph{c}-axis, showing that
the two components are of comparable magnitude, resulting in full
directional sensitivity to arbitrary electric fields. This is a unique
property that distinguishes V\textsubscript{Si} from other spin defects
for quantum electrometry. Furthermore, we prove that the electric dipole
moments remain unchanged even in the high electric field region of
$\sim$2.3 MV/cm, approaching 90\% of the breakdown electric
field of the measured sample. Selectively formed V\textsubscript{Si}
dots yield high-resolution three-dimensional maps of the electric field
distribution. These results confirm that V\textsubscript{Si}-based
quantum electrometry can probe the full electric field regime
encountered in SiC power devices with nanometre-scale spatial precision.

\begin{figure}[tb]
\centering{\includegraphics[width=8cm]{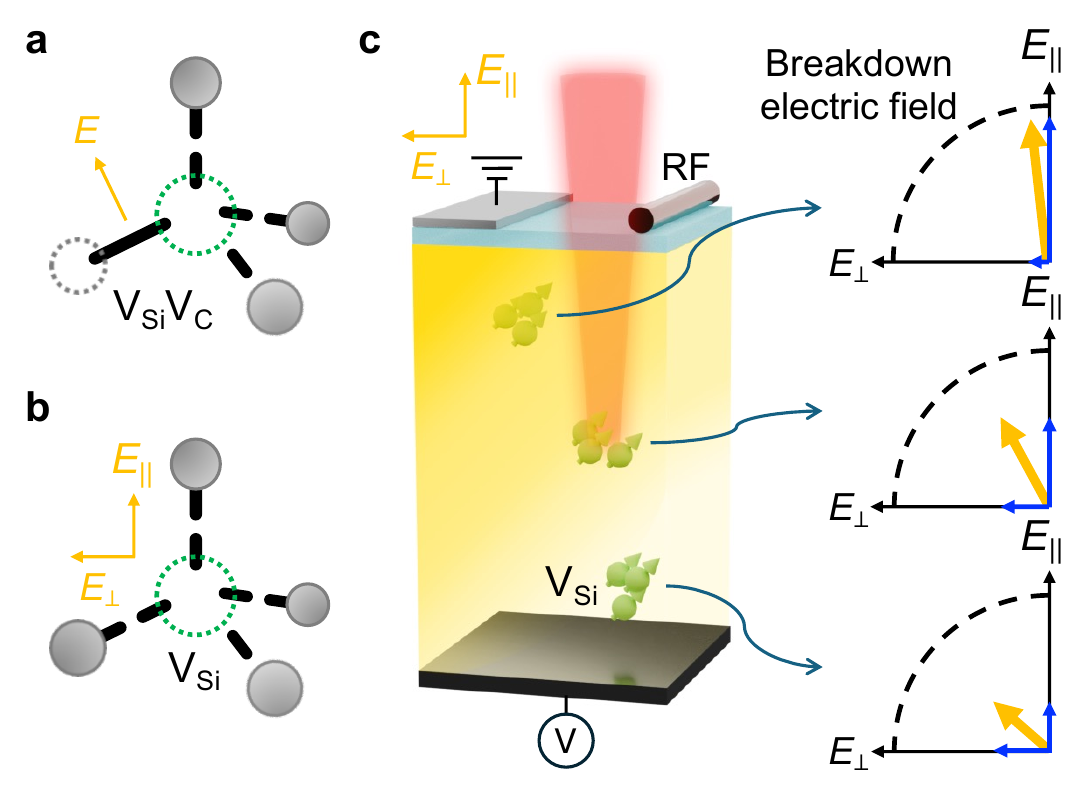}}
\caption{Conceptual diagram of quantum electrometry in a SiC power
device.
\textbf{a}, Structure of V\textsubscript{Si}V\textsubscript{C} in
4H-SiC. This defect responds well to an electric field parallel to its
defect axis. \textbf{b}, Structure of V\textsubscript{Si} in 4H-SiC.
This defect, being a monovacancy with no defect axis, responds equally
to electric fields applied both parallel and perpendicular to the
quantization axis of V\textsubscript{Si} (= \emph{c}-axis). \textbf{c},
An electric field inside a SiC device is measured with high spatial
resolution using a selectively formed V\textsubscript{Si}-based quantum
sensors embedded within the device. Selective formation of
V\textsubscript{Si} avoids device degradation and enables electric field
measurement while maintaining the device in a normal state. A red laser
focused on one of ensemble V\textsubscript{Si} and a metal wire for
applying RF for quantum sensing are depicted.}
\end{figure}

\section*{Determination of electric dipole moments}
Figure 2a shows the structure of the pn diode with an edge termination
region used in this study (see Methods for details). To apply reverse
bias, a positive voltage was applied to the bottom electrode, and the
top electrode was connected to GND. In the absence of any device
anomaly, an internal electric field can be reliably reproduced by
technology computer-aided design (TCAD) device simulations (see
Supplementary Information 1). For our device structure, an electric field,
\textbf{E} = (\emph{E\textsubscript{x}}, \emph{E\textsubscript{y}},
\emph{E\textsubscript{z}}), was directed almost along the surface normal
(\emph{z} direction) as indicated by the red arrows in Fig. 2a.
E\emph{\textsubscript{x}} is one order of magnitude smaller than
\emph{E\textsubscript{z}} (\emph{E\textsubscript{y}} = 0 at all
measurement positions). Because the device is fabricated on a n-type
substrate with a 4$^\circ$-off orientation in the {[}11$\bar{2}$0{]} direction, the
distributions of \emph{E}\textsubscript{z} and \emph{E}\textsubscript{x}
calculated by TCAD simulations were corrected to obtain the
distributions of the electric field components parallel
(\emph{E}\textsubscript{\textbar\textbar{}}) and perpendicular
(\emph{E}\textsubscript{$\bot$}) to the \emph{c}-axis. Figures 2b,c show the
simulated distributions of \emph{E}\textsubscript{\textbar\textbar{}},
whose maximum value reaches $\sim$2.3 MV/cm at an applied
voltage of 1500 V. Figures 2d,e show the simulated distributions of
\emph{E}\textsubscript{$\bot$}. Due to the 4$^\circ$ tilt of the \emph{c}-axis,
\emph{E}\textsubscript{$\bot$} is almost zero near point1 in Fig. 2d and
takes a large value near point3 in Fig. 2e.

\begin{figure}[tb]
\centering{\includegraphics[width=14cm]{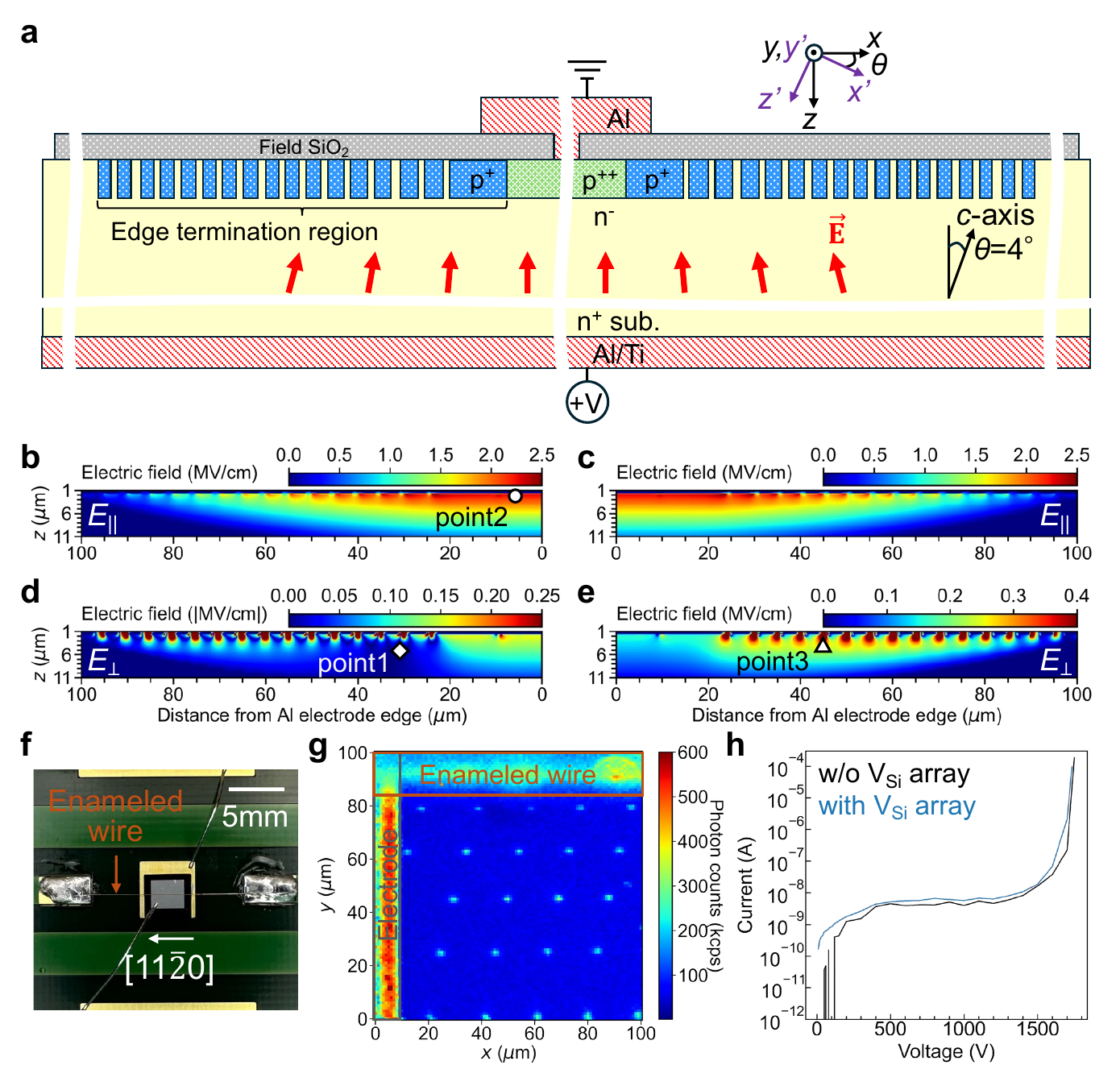}}
\caption{Sample structure and electric properties.
\textbf{a}, Cross-sectional image of a pn diode with an edge termination
region. The device is fabricated on a n-type substrate with a 4$^\circ$-off
orientation. \textbf{b-e}, Distribution of electric field components
parallel (\emph{E}\textsubscript{\textbar\textbar{}}) and perpendicular
(\emph{E}\textsubscript{$\perp$}) to the \emph{c}-axis obtained by correcting
device simulation results. Applied voltage is set to 1500 V. Point1 and
point3 are used to determine the values of
\emph{d}\textsubscript{\textbar\textbar{}} and
\emph{d}\textsubscript{$\perp$}, respectively, and at point2, high electric
field measurement is performed to investigate the stability of
\emph{d}\textsubscript{\textbar\textbar{}} and \emph{d}\textsubscript{$\bot$}
under high electric field conditions. \textbf{f}, Top view of a sample
mounted on a PCB. The upper electrode (p-type epilayer) is connected to
GND, while a positive voltage is applied to the lower electrode (n-type
substrate). To avoid a short circuit between the RF and DC lines, an
enameled wire is used for the RF application. The white arrow indicates
the off-direction ({[}11$\bar{2}$0{]}). \textbf{g}, Photoluminescence mapping
image of the V\textsubscript{Si} dot array. \textbf{h}, I-V
characteristics under reverse bias before and after V\textsubscript{Si}
formation. No significant changes were observed after PBW.}
\end{figure}

A sample was mounted on a printed circuit board (PCB) with DC and RF
lines as shown in Fig. 2f, and the whole was immersed in Fluorinert (see
Methods and Supplementary Information 2 for details of the setup). To probe an
internal electric field, we introduced a V\textsubscript{Si} (ensemble)
dot array using the particle beam writing (PBW) method \cite{a28, a29}.
Each bright spot in Fig. 2g corresponds to a V\textsubscript{Si}
ensemble and the V\textsubscript{Si} density within each dot was set to
5$\times$10\textsuperscript{15} cm\textsuperscript{-3} (see Supplementary
Information 3). Figure 2h shows current-voltage (I-V) characteristics under
reverse bias before and after the formation of the V\textsubscript{Si}
dot array. The black and blue curves in Fig. 2h are essentially
identical, indicating that the PBW process does not degrade device
performance (see Supplementary Information 4).

Firstly, we determined the electric dipole moment parallel to the
\emph{c}-axis (\emph{d}\textsubscript{\textbar\textbar{}}). Point1
(diamond mark in Fig. 2d) is the position where the maximum
\emph{E}\textsubscript{\textbar\textbar{}} is obtained within the region
where the ratio of \emph{E}\textsubscript{$\bot$} to
\emph{E}\textsubscript{\textbar\textbar{}}
(\emph{E}\textsubscript{$\bot$}/\emph{E}\textsubscript{\textbar\textbar{}})
is sufficiently small (\textless{} 0.01). As shown later, a large
\emph{E}\textsubscript{$\bot$}/\emph{E}\textsubscript{\textbar\textbar{}}
introduces nonlinearity into the change in resonance frequency.
Therefore, selecting the point1 enables the most accurate determination
of \emph{d}\textsubscript{\textbar\textbar{}} in the sample. Figure 3a
shows ODMR spectra obtained at the applied voltages ranging from 0 to
1000 V. The black lines are the fitting curves. At zero applied voltage,
the resonant frequency was determined to be $\sim$71 MHz, which
is consistent with 2\emph{D}, where \emph{D} is the zero-field-splitting
parameter for the ground state of V\textsubscript{Si}, $\sim$35
MHz \cite{a30}. As the applied voltage increased, the resonance
frequencies shifted toward lower frequencies. Figure 3b shows the
resonant frequencies as a function of a simulated electric field value
(\emph{E}\textsubscript{\textbar\textbar{}}) at point1. The resonant
frequency decreased linearly with increasing
\emph{E}\textsubscript{\textbar\textbar{}}, which is consistent with the
change in resonance frequency with respect to
\emph{E}\textsubscript{\textbar\textbar{}},
\(f = \ \left| 2\left( D + \frac{d_{||}}{h}E_{||} \right) \right|\),
predicted by theoretical analysis (see Methods). From the result,
\emph{d}\textsubscript{\textbar\textbar{}}/\emph{h} was determined to be
-15.0 ± 0.5 MHz/(MV/cm). This value is close to that of our theoretical
calculations (-15.6 MHz/(MV/cm)) (see Supplementary Information 5) and the
reported theoretical calculation values \cite{a31}. The reported
experimental value (-3.95 MHz/(MV/cm)) in the previous study \cite{a23} is
four times smaller than our value because the previous work defined the
electric field as the sum of the macroscopic and the local dipole
fields. Using only the macroscopic field, their result becomes
consistent with our value.

\begin{figure}[tb]
\centering{\includegraphics[width=16cm]{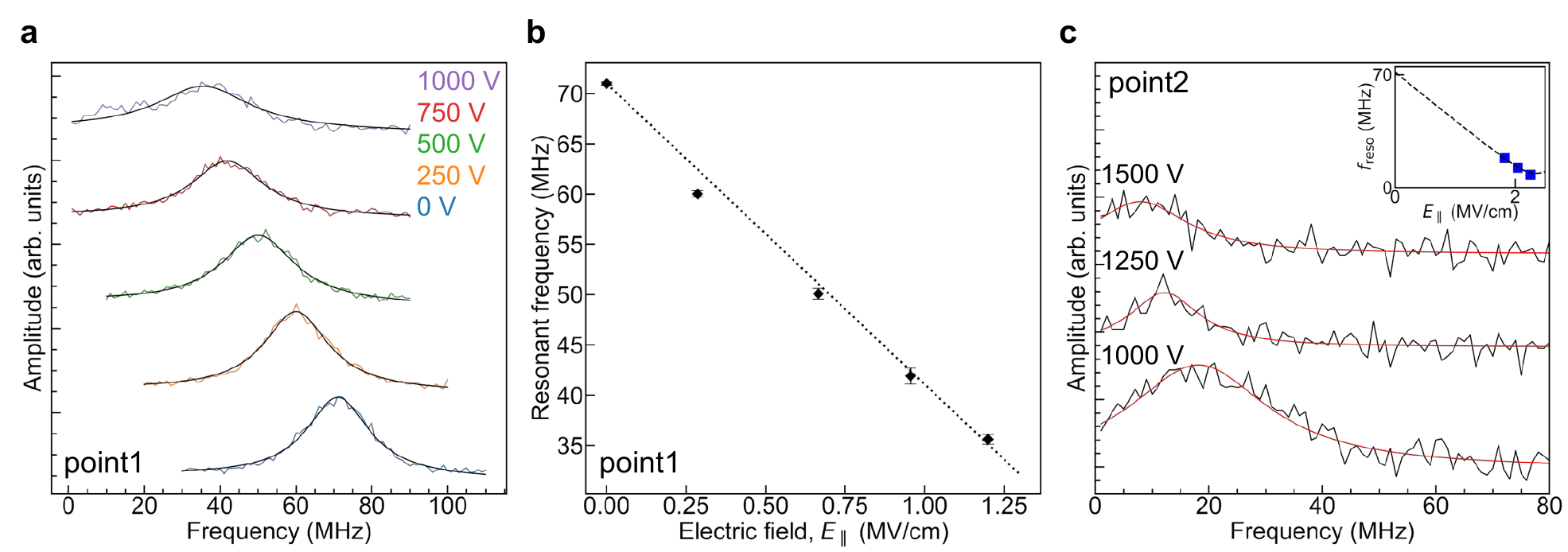}}
\caption{Determination of \emph{d}\textsubscript{\textbar\textbar{}}
and its stability under high electric fields.
\textbf{a}, ODMR spectra at point1 (diamond mark in fig. 2d) measured
under reverse bias ranging from 0 to 1000 V. The black lines are the
fitting curves. \textbf{b}, Electric field dependence of the resonance
frequency. The values of electric field are based on device simulations.
From the gradient, \emph{d}\textsubscript{\textbar\textbar{}}/\emph{h}
was determined to be -15.0 ± 0.5 MHz/(MV/cm). \textbf{c}, ODMR spectra
at point2 (circle mark in fig. 2b) measured under applied voltages of
1000, 1250 and 1500 V. Electric fields based on device simulation are
$\sim$1.8, $\sim$2.1 and $\sim$2.3 MV/cm,
respectively. Inset shows the theoretical values (dashed line)
considering related parameters
(\emph{d}\textsubscript{\textbar\textbar{}},
\emph{E}\textsubscript{$\perp$}/\emph{E}\textsubscript{\textbar\textbar{}} at
the measured point and
\textbar{}\emph{d}\textsubscript{$\perp$}/\emph{d}\textsubscript{\textbar\textbar{}}\textbar)
and the experimental values. It is noted that even at an
\emph{E}\textsubscript{$\perp$}/\emph{E}\textsubscript{\textbar\textbar{}}
ratio of 0.07, nonlinearity in the resonance frequency shift appears,
especially at high electric fields exceeding 2 MV/cm.}
\end{figure}

Next, we show the experimental results under high electric fields near
the breakdown electric field of SiC. The ODMR spectra were measured at
point2 marked with a circle in Fig. 2b, where the highest electric field
of $\sim$2.3 MV/cm is obtained at 1500 V, while the maximum
electric field at point1 is only $\sim$1.2 MV/cm. Figure 3c
shows the ODMR spectra under applied voltages from 1000 to 1500 V. These
results show that the V\textsubscript{Si}-based quantum sensor can
measure an electric field of up to $\sim$2.3 MV/cm. The
measurable electric field reaches $\sim$90\% of the breakdown
electric field which is calculated to be $\sim$2.5 MV/cm using
the doping concentration of 1$\times$10\textsuperscript{16}
cm\textsuperscript{-3} for n-type epilayer \cite{a32}. In other words,
V\textsubscript{Si} operates as a quantum sensor even under strong
electric fields approaching the material's physical limit. This further
indicates that the introduction of V\textsubscript{Si} does not
adversely affect the performance of the host device. The inset of Fig.
3c shows experimental resonant frequencies and theoretical resonant
frequency shift. The latter was calculated considering the value of
\emph{d}\textsubscript{\textbar\textbar{}},
\emph{E}\textsubscript{$\bot$}/\emph{E}\textsubscript{\textbar\textbar{}}
($\sim$0.07) at point2 and
\textbar{}\emph{d}\textsubscript{$\bot$}/\emph{d}\textsubscript{\textbar\textbar{}}\textbar{}
(= $\sim$1.1) which will be discussed later. The experimental
data agrees well with the theoretical values, indicating that the values
of \emph{d}\textsubscript{\textbar\textbar{}} and
\emph{d}\textsubscript{$\bot$} remain unchanged under the high electric
fields.

The electric dipole moment perpendicular to the \emph{c}-axis,
\emph{d}\textsubscript{$\bot$}, was also investigated in our sample. When the
applied electric field has not only the
\emph{E}\textsubscript{\textbar\textbar{}} component but also the
\emph{E}\textsubscript{$\bot$} component, an additional modulation that
depends on \emph{d}\textsubscript{$\bot$} appears in the resonant frequency
shift. The calculated resonance frequencies for different
\emph{d}\textsubscript{$\bot$}/\emph{d}\textsubscript{\textbar\textbar{}} are
shown as solid lines in Fig. 4. The value of \emph{d}\textsubscript{$\bot$}
can be experimentally estimated from the \emph{d}\textsubscript{$\bot$}
dependence of the resonant frequency shift. For this purpose, locations
where both \emph{E}\textsubscript{$\bot$} and
\emph{E}\textsubscript{$\bot$}/\emph{E}\textsubscript{\textbar\textbar{}} are
large are suitable. In addition, as shown in Fig. 4, a higher
\emph{E}\textsubscript{\textbar\textbar{}}, for example more than 1
MV/cm at higher applied voltages, is also required to realize a
sufficiently large additional modulation in the resonant frequency shift
to obtain a more accurate value of \emph{d}\textsubscript{$\bot$}.
Considering these conditions, we selected the point3 (triangle mark in
Fig. 2e).
\emph{E}\textsubscript{$\bot$}/\emph{E}\textsubscript{\textbar\textbar{}} at
point3 is $\sim$0.19, and the maximum electric field of
$\sim$1.5 MV/cm is applied under applied voltage of 1500 V.
Comparing the experimental results (triangle marks in Fig. 4) and
calculated values (solid lines in Fig. 4),
\textbar{}\emph{d}\textsubscript{$\bot$}/\emph{d}\textsubscript{\textbar\textbar{}}\textbar{}
is determined to be = 1.1 ± 0.06. Our theoretical calculations yield
\emph{d}\textsubscript{$\bot$} $\sim$ +16 MHz/(MV/cm), resulting in
\textbar{}\emph{d}\textsubscript{$\bot$}/\emph{d}\textsubscript{\textbar\textbar{}}\textbar{}
$\sim$ 1.0. The experimental value is in close agreement with
the theoretically predicted value. The positive sign of the theoretical
value for \emph{d}\textsubscript{$\bot$} matches the theoretical calculations
performed for pressure which has the same effect on the spin Hamiltonian
as an electric field \cite{a31}. It is noted that the sign cannot be
determined using the method employed in this study because the resonance
frequency shift does not depend on the direction of
\emph{E}\textsubscript{$\bot$}.

\begin{figure}[tb]
\centering{\includegraphics[width=8cm]{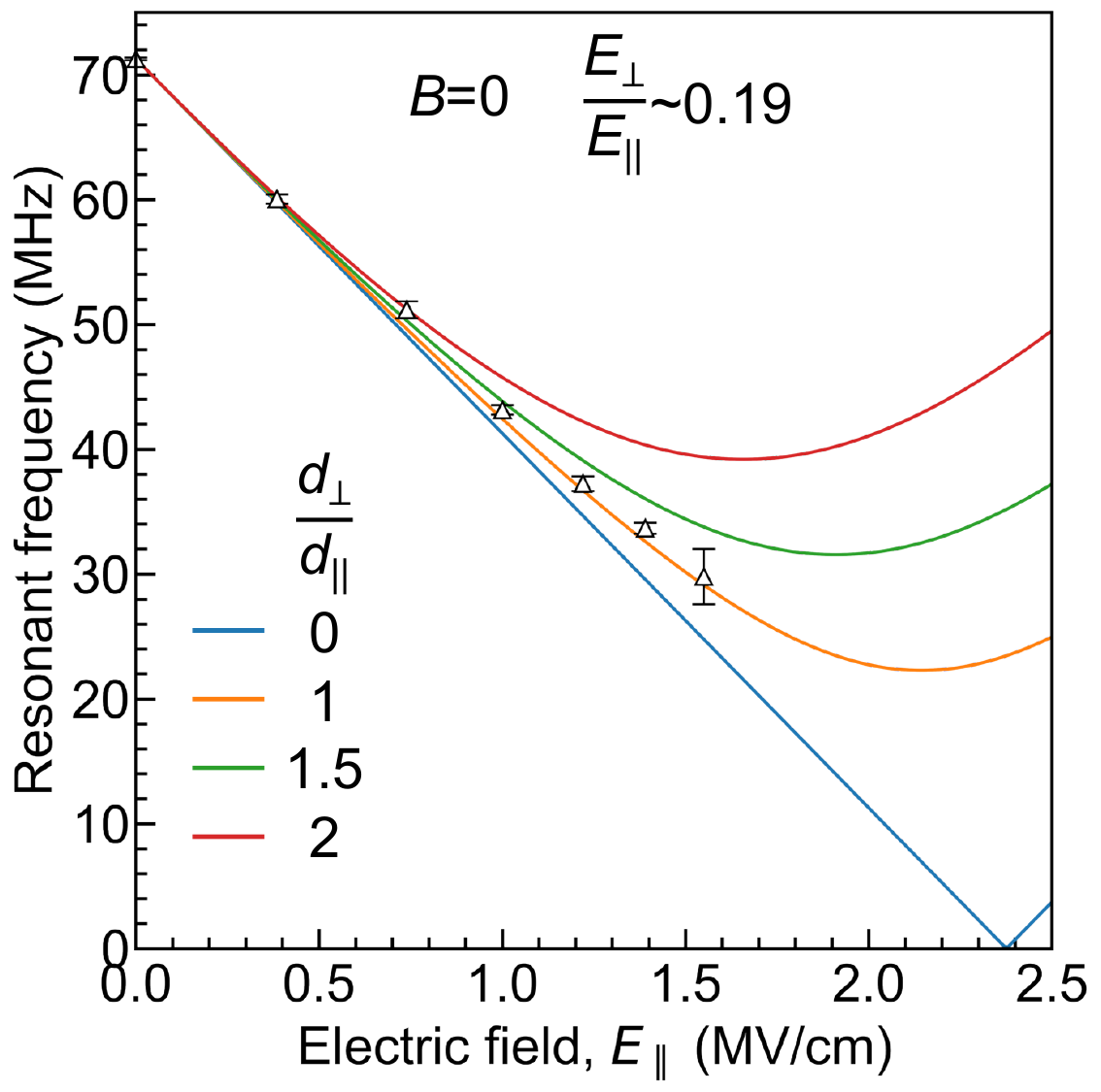}}
\caption{Determination of \emph{d}\textsubscript{$\bot$}.
Electric field dependence of the resonance frequency at point3 (triangle
mark in fig. 2e) and the theoretical resonance frequencies calculated by
varying
\emph{d}\textsubscript{$\bot$}/\emph{d}\textsubscript{\textbar\textbar{}}
from 0 to 2 under a zero magnetic field.
\emph{E}\textsubscript{$\bot$}/\emph{E}\textsubscript{\textbar\textbar{}} was
calculated to be $\sim$0.19 at the point. Comparing the
experimental data and calculated values,
\textbar{}\emph{d}\textsubscript{$\bot$}/\emph{d}\textsubscript{\textbar\textbar{}}\textbar{}
is estimated to be 1.1 ± 0.06.}
\end{figure}

Table 1 summarizes the electric dipole moments of various spin defects
\cite{a22}. For V\textsubscript{Si},
\emph{d}\textsubscript{\textbar\textbar{}} and \emph{d}\textsubscript{$\bot$}
have values of the same order, which is significantly different from the
others. This is presumed to be due to the structural feature of a
monovacancy, which is free from strong uniaxial anisotropy such as those
of V\textsubscript{Si}V\textsubscript{C} in SiC and the nitrogen-vacancy
(NV) center in diamond. In addition, the values are also relatively
large compared to the other spin defects, and remain unchanged even with
an electric field up to $\sim$2.3 MV/cm. These outstanding
characteristics of the electric dipole moments is practical for device
evaluation in structures such as trench MOS gates and super-junction
structures \cite{a25, a26, a27}, where control of the lateral electric field
(\emph{E}\textsubscript{$\bot$}) is a critical issue.

\begin{table}[tb]
 \caption{Electric dipole moments in the ground state of various spin
defects.
The values for the NV center in diamond and PL1-6 in SiC are from Ref. \cite{a22}. PL1-4 are V\textsubscript{Si}V\textsubscript{C}. PL5 and PL6
are identified as a near-stacking fault axial
V\textsubscript{Si}V\textsubscript{C} \cite{a33}. For the NV center and
V\textsubscript{Si}V\textsubscript{C},
\emph{d}\textsubscript{\textbar\textbar{}} and \emph{d}\textsubscript{$\bot$}
represent electric dipole moments parallel and perpendicular to the
defect axis, respectively. The unit is Hz/(V/cm).
}
 \centering
  \begin{tabular}{|c|c|c|c|c|c|c|c|c|}
   \hline
    & V\textsubscript{Si} & NV center & PL1(\emph{hh}) & PL2(\emph{kk}) & PL3(\emph{hk}) & PL4(\emph{kh}) & PL5 & PL6\\
   \hline
   \textbar \emph{d}\textsubscript{\textbar\textbar{}}/\emph{h}\textbar{} & 15.0 & 0.35 & 2.65 & 1.61 & \textless3 & 0.44 & \textless3 & 0.96 \\
   \hline
   \textbar \emph{d}\textsubscript{$\bot$}/\emph{h}\textbar{} & 16.5 & 17 & & & 32.3 & 28.5 & 32.5 &\\
   \hline
  \end{tabular}
\end{table}

We also discuss resonance frequency correction due to temperature
changes. Temperature variations cause shifts in the resonance frequency
just as an electric field. Therefore, to accurately determine an
electric field, it is necessary to correct for the resonance frequency
changes caused by temperature variations that may occur inside a device
in operation. Fortunately, for V\textsubscript{Si}, \emph{D} remains
unchanged over a wide temperature range of 300-590K \cite{a24}. This
unique property of \emph{D} is not observed in other spin defects
\cite{a34, a35, a36}, and offers a practical advantage: it enables accurate
electric field measurements without requiring correction for
temperature-induced changes in resonance frequency even near areas where
current concentrations cause abnormal heating.

\section*{Electrical field distribution measurement}

High-resolution mapping of an electric field inside an operating device
enables the diagnosis of various failure mechanisms, leading to improved
device reliability. We therefore measured the electric field
distribution inside the device. The V\textsubscript{Si} dot depth was
varied from 2.1 to 8.1 $\mu$m by changing the ion energy in the PBW from
0.75 to 3 MeV, where the depths of V\textsubscript{Si} are based on the
values predicted by stopping and range of ions in matter (SRIM)
simulation {[}http://www.srim.org/ for ``SRIM: The Stopping and Range of
Ions in Matter.''{]} (see Supplementary Information 6). Figure 5a summarizes
the results. For \emph{x} \textless{} 50 $\mu$m, the experimental results
(circle marks) agreed well with the values obtained from device
simulations (dashed lines). However, significant discrepancies were
observed for \emph{x} \textgreater{} 50 $\mu$m regions. The discrepancies
decreased as the depth of V\textsubscript{Si} increased. The
discrepancies are partly owing to the device simulation not accounting
for the influence of both the enamel wire with DC potential of zero and
dielectric constant of Fluorinert on the electric field distribution.
Figures 5b,c show the electric field distributions obtained by device
simulations with and without the metal conductor (assuming the enameled
wire) and adjusted dielectric coefficient between the metal conductor
and the sample surface (see Supplementary Information 1). The electric field
was negligibly small at \emph{x} \textgreater{} 90 $\mu$m without the
enameled wire (dashed lines in Fig. 5a). In contrast, device simulations
with a parallel metal conductor placed 4 $\mu$m from the surface revealed
that at the surface side (e.g., solid line for \emph{z} = 3.7 $\mu$m in Fig.
5a), the electric field extends to up to \emph{x} $\sim$ 100
$\mu$m. The tendency of experimental values agreed with device simulations
considering the metal conductor and dielectric constant of Fluorinert
(solid lines in Fig. 5a). Since the discrepancies of electric field
distribution is mainly caused by a foreign material on the sample, the
influence is expected to decrease as the distance between the foreign
material and the device increases. This is confirmed by other device
simulations (see Supplementary Information 1). Therefore, the discrepancies can
be easily avoided across the entire region by setting the antennas used
for RF application far enough away from the sample. The results
demonstrate that a V\textsubscript{Si}-based quantum sensor can
accurately measure electric field in the device with high spatial
resolution, including disturbances in the electric field distribution.

\begin{figure}[tb]
\centering{\includegraphics[width=10cm]{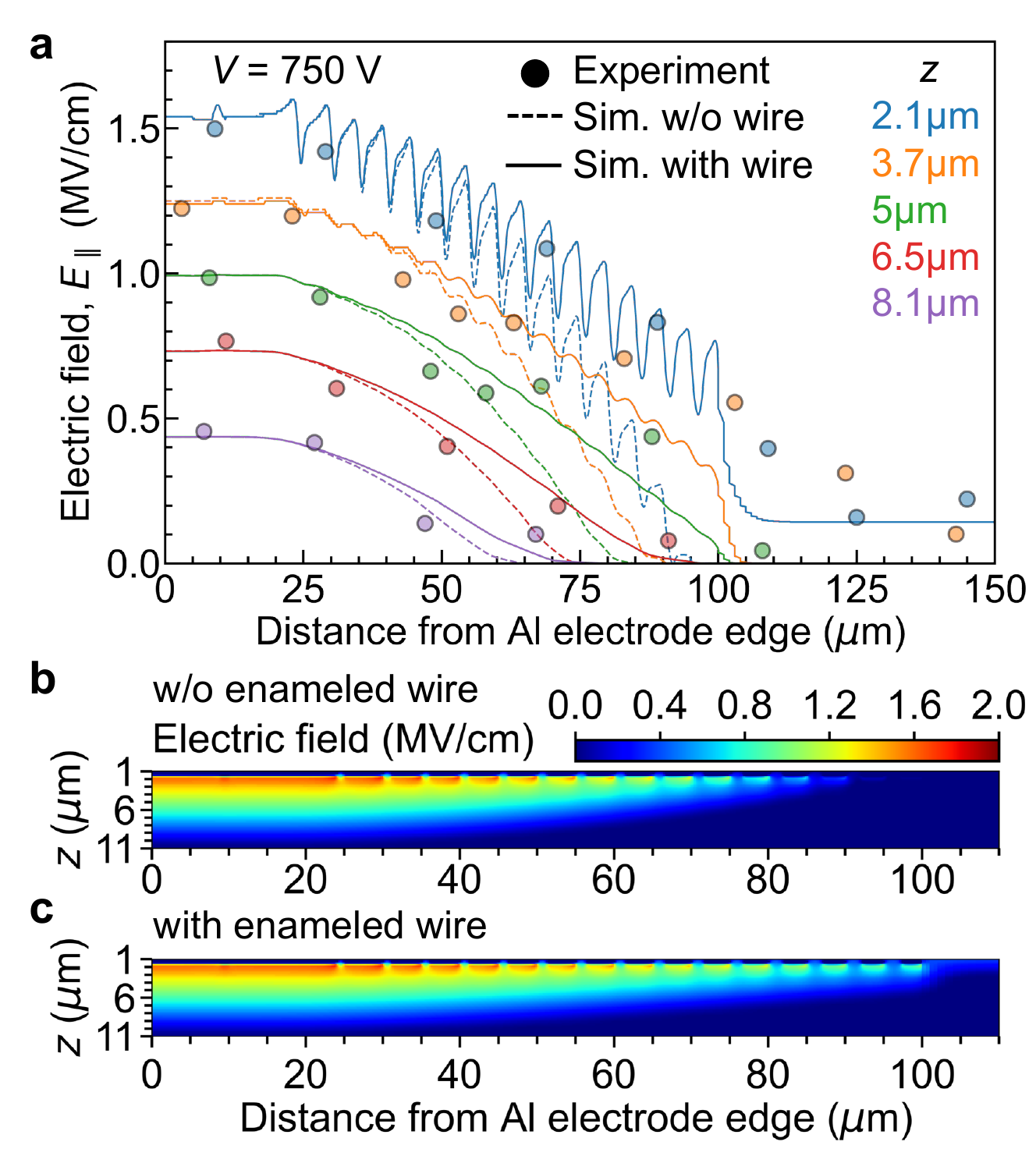}}
\caption{Electrical field distribution measurements.
\textbf{a}, Electric field distribution measured under the applied
voltage of 750 V. Electric fields were calculated from
\emph{d}\textsubscript{\textbar\textbar{}} and the resonance
frequencies. Since the effect of \emph{E}\textsubscript{$\bot$} on the
resonance frequency is small under the condition, no correction
accounting for \emph{E}\textsubscript{$\bot$} was applied at each measurement
point. Solid and dashed lines show the simulated electric fields with
and without the enameled wire with DC potential of zero, respectively.
The discrepancies from the dashed line are significant in the regions
\emph{z} = 3.7 and 5  $\mu$m, and \emph{x} \textgreater{} 70  $\mu$m. \textbf{b},
\textbf{c}, Electric field distribution obtained from device simulations
without and with the enameled wire, respectively. The presence of the
enameled wire causes electric field to extend to \emph{x} \textgreater{}
100  $\mu$m.}
\end{figure}

This feature makes electric field detection technique using
V\textsubscript{Si}-based quantum sensors promising not only for
analyzing normal devices but also for analyzing degraded devices, such
as those exhibiting crystalline degradation due to long-term operation.
The latter is particularly difficult to reproduce in device simulations,
suggesting that V\textsubscript{Si}-based quantum sensors could be a
particularly effective tool for evaluating failure mechanisms of SiC
power devices. For example, this could be applied to device health
assessment. Defects near the SiO\textsubscript{2}/SiC interface and a
single Shockley-type stacking fault generated by basal plane
dislocations during device operation remain a critical issue \cite{a37},
and solutions are currently desired \cite{a38, a39}. It may also be
possible to detect external stray electric fields. Although verification
at depths exceeding 8 $\mu$m was not possible due to the limitations of the
sample structure and the available ion energy in the PBW, we have
reported that ODMR measurements are possible up to the maximum depth of
$\sim$60 $\mu$m using V\textsubscript{Si} dot array fabricated by
the PBW with H ions \cite{a40}. Selective formation technique capable of
forming V\textsubscript{Si} to a depth of several tens of micrometers
can cover all regions requiring failure analysis for typical SiC power
devices, such as critical structures in deep regions in a trench gate
MOSFET and super-junction device.

\section*{Conclusions}

We showed that V\textsubscript{Si} formed by the PBW can be used as
quantum sensors to measure electric fields inside high-biased SiC power
devices with high spatial resolution in operando. Electric dipole
moments (\emph{d}\textsubscript{\textbar\textbar{}} and
\emph{d}\textsubscript{$\bot$}) were experimentally determined up to high
electric fields approaching the material's physical limit. Unlike other
spin defects, the values of \emph{d}\textsubscript{\textbar\textbar{}}
and \emph{d}\textsubscript{$\bot$} were quite close, which is presumed to be
due to the structural feature of a monovacancy. This excellent
characteristic is of practical importance because V\textsubscript{Si}
can be used to measure electric fields applied in various directions
with equivalent sensitivity. V\textsubscript{Si}-based quantum sensors
can measure not only electric field up to $\sim$2.3 MV/cm,
sufficiently high for practical applications, but also electrical field
distribution. The latter also indicates the potential to detect
disturbances in electric field caused by external artifacts and/or stray
electric fields. Combining the characteristic electric dipole moments
with high spatial resolution given by selective formation technique,
V\textsubscript{Si}-based quantum sensor is anticipated to function as
an effective tool for elucidating electric field distributions in
various types of SiC power devices, which has been difficult to achieve
until now.

From a practical point of view, however, it must be noted that for
electric field measurements, there is another limitation concerning the
number of V\textsubscript{Si} which is a key parameter determining
sensor sensitivity as shown in Supplementary Information 3. This limitation is
not present in magnetic field and temperature measurements.

The operando electric field measurements by V\textsubscript{Si}-based
quantum sensors not only opens a path to data-driven research and
development to improve device performance, but also for a broad range of
applications with potential as a fundamental research tool for device
health assessment and studies of defects, dislocations, and other
phenomena.

\section*{Methods}

\section*{Sample preparation}

A pn diode with an edge termination region was fabricated on a 4$^\circ$-off
n-type SiC substrate. Doping concentrations and thickness of n-type
epilayer are 1$\times$10\textsuperscript{16} cm\textsuperscript{-3} and 12
$\mu$m, respectively. By applying reverse bias below the breakdown volage of
the diode ($\sim$1500 V), a high electric field up to
$\sim$2.3 MV/cm can be generated. V\textsubscript{Si}
(ensemble) dot array was formed by PBW using a focused beam of He ions
with a diameter of 1 $\mu$m. The irradiation fluence was set to 5$\times$10\textsuperscript{3} ions/spot ($\sim$ 5$\times$10\textsuperscript{11} cm\textsuperscript{-2}) to achieve the
V\textsubscript{Si} density within each dot of $\sim$ 5$\times$10\textsuperscript{15} cm\textsuperscript{-3}. The reason for this is
that accurate electric field measurement is not possible at higher
V\textsubscript{Si} densities (see Supplementary Information 3). An ion energy
was varied from 0.75 to 3 MeV to control the depth of
V\textsubscript{Si}. To minimizes the total irradiation fluence,
V\textsubscript{Si} dots with 1 $\mu$m in diameter were formed at 20 $\mu$m
intervals. This prevents device degradation caused by introduction of
various defects. Detailed I-V data before and after V\textsubscript{Si}
formation are shown in Supplementary Information 4. The leakage current is
maintained below approximately 10 nA when a maximum voltage of
$\sim$1500 V was applied. At the current levels, ODMR
measurements are not expected to be affected at all \cite{a41}.

\section*{TCAD device simulation}

We used commercial TCAD device simulator (Synopsys Sentaurus Version
X-2025.06), which provides a comprehensive modeling capabilities for
semiconductor device physics. The electric field simulations solve
Poisson's equation coupled to carrier transport (drift-diffusion or
hydrodynamic) and the continuity equations, incorporating material
properties and recombination-generation physics. Specifically, the
internal electric field in the edge-termination region was computed in
the \emph{SDevice} module using the Bank-Rose nonlinear solver with the
ILS method. Material parameters were adopted from the references \cite{a1, a3}, and a p-type doping profile measured by Secondary Ion Mass
Spectrometry was incorporated into the models. Two device configurations
were modeled: (1) a standalone device in isolation, (2) a device
adjacent to a conductive wire biased at 0 V. Details are provided in
Supplementary Information 1.

\section*{Confocal microscope (CFM) setup}

Measurements were performed using a home-made CFM. The wavelength of the
excitation laser was 785 nm, and photons were detected with an avalanche
photodiode after passing through a 900 nm long pass filter. To avoid
discharge due to the application of high voltages, the entire PCB board
on which the sample was set (including DC and RF lines) was immersed in
Fluorinert. As shown in Supplementary Information 2, since Fluorinert exhibits
a transmittance of approximately 95\% with respect to the excitation
laser and V\textsubscript{Si} photoemission wavelengths, the immersion
does not affect measurements. Considering the refractive index of
Fluorinert (= 1.29), a water immersion objective lens was chosen to
improve photon collection efficiency. RF was amplified and then applied
using the enameled wire. All measurements were performed at room
temperature.

\section*{Theoretical analysis of the change in the resonant frequency under
\emph{E}\textsubscript{\textbar\textbar{}}}

Similar to the NV center in diamond, V\textsubscript{Si} has the
\emph{C\textsubscript{3v}} symmetry \cite{a42}. Therefore, the ground
state spin Hamiltonian, \emph{H}, including the electric dipole moments
(\emph{d}\textsubscript{\textbar\textbar{}} and
\emph{d}\textsubscript{$\bot$}), can be written in
\emph{x'}-\emph{y}-\emph{z'} coordinate, where the
\emph{z\textquotesingle{}}-axis is parallel to the \emph{c}-axis
\cite{a19},

\(H = g\mu_{B}\mathbf{S \cdot B} + \left( hD + d_{||}E_{z'} \right)\left\lbrack S_{z'}^{2} - \frac{S(S + 1)}{3} \right\rbrack - d_{\bot}\left\lbrack E_{x'}\left( S_{x'}S_{y'} + S_{y'}S_{x'} \right) + E_{y'}\left( S_{x'}^{2} - S_{y'}^{2} \right) \right\rbrack\),

where \(g\) is the electron g-factor, $\mu$\textsubscript{B} is the
Bohr magneton, \textbf{S} is the electron spin operator with the spin
quantum number S = 3/2, \textbf{B} is the applied magnetic field,
\emph{h} is the Planck constant, and \emph{D} is the
zero-field-splitting parameter for the ground state of
V\textsubscript{Si}. Solving this under the conditions
\textbar{}\textbf{B}\textbar{} = \emph{B\textsubscript{z'}}
(\emph{B\textsubscript{x'}} = \emph{B\textsubscript{y'}} = 0) and
\textbar{}\textbf{E}\textbar{} = \emph{E\textsubscript{z'}}
(\emph{E\textsubscript{x'}} = \emph{E\textsubscript{y'}} = 0) yields
resonance frequencies
\(f_{\pm} = \ \left| 2\left( D + \frac{d_{||}}{h}E_{z'} \right) \pm \frac{g\mu_{B}}{h}B_{z'} \right|\).
From this formula, we can see that when only
\emph{E}\textsubscript{\textbar\textbar{}} (=
\emph{E\textsubscript{z'}}) is applied, the resonant frequency shifts
linearly with respect to \emph{E}\textsubscript{\textbar\textbar{}}.

\section*{Theoretical calculation for electric dipole moments}

We calculated \emph{d}\textsubscript{\textbar\textbar{}}/\emph{h} and
\emph{d}\textsubscript{$\bot$}/\emph{h} using the first-principles spin
density functional calculations. The calculations were done with the
VASP package using the projector-augmented wave method for atomic
pseudopotentials \cite{a43, a44, a45, a46, a47}. The GGA-PBE functional \cite{a48} was taken
for approximating the exchange-correlation energy. Energy cutoff for the
plane waves was set to 500 eV. Only Gamma point was considered for the
self-consistent charge density and D-tensor (second order coefficients
of the effective spin Hamiltonian) calculations. The latter was
calculated using the Rayson-Briddon formula \cite{a49} as implemented in
VASP. Comparing the calculated D-tensor under varying macroscopic fields
at the target V\textsubscript{Si}\textsuperscript{-} defect, we
estimated \emph{d}\textsubscript{\textbar\textbar{}}/\emph{h} and
\emph{d}\textsubscript{$\bot$}/\emph{h} as the linear-response coefficients.

To characterize the internal macroscopic field that corresponds to the
experimental conditions, we made a supercell slab of 4H-SiC in a vacuum,
which include 5$\times$5$\times$2 units of (SiC)\textsubscript{4} and appropriate
surface radicals to remove artificial dangling bonds from the bulk gap
region. Charged V\textsubscript{Si}\textsuperscript{-} at a k-site was
formed by removing one Si atom and replacing another Si at a distant
site on the same plane with P atom (P\textsubscript{Si}). The
first-principles calculation yields the effective electrostatic
potential induced by the target defect itself. To estimate the
macroscopic field exerted by the other sources, the component parallel
to \emph{z'} direction, \emph{E\textsubscript{z\textquotesingle{}}}, was
estimated by analyzing the planar-averaged electrostatic potential for
the defect-free cases calculated with VASPKIT \cite{a50}, whereas
\emph{E\textsubscript{x\textquotesingle{}}} and
\emph{E\textsubscript{y\textquotesingle{}}} were estimated by
approximating V\textsubscript{Si}\textsuperscript{-},
P\textsubscript{Si}\textsuperscript{+} and their translational images as
point charges placed within a uniform medium with dielectric constant
10.0 ($\approx$experimental value of the bulk SiC). See Supplementary Information 5 for
more details.

\section*{Data availability}
The data that support the findings of this study are available from the
corresponding author upon reasonable request.

\section*{Acknowledgements}
This work was supported by the Council for Science, Technology and
Innovation (CSTI), Cross-ministerial Strategic Innovation Promotion
Program (SIP), under the research theme ``Environment development for
practical use of solid-state quantum sensors: towards social
implementation,'' within the project ``Promoting application of advanced
quantum technologies to social challenges'' (Funding agency: QST). This
study is partially supported by MEXT Quantum Leap Flagship Program (MEXT
Q-LEAP) Grant No. JPMXS0118067395. This work was partially supported by
JSPS KAKENHI under Grant-in-Aid for Early-Career Scientists No.
JP24K16985 and JSPS KAKENHI under Grant-in-Aid for Transformative
Research Areas No. JP22H05114. We thank Professor A. Gali from Budapest
University of Technology and Economics, Hungary for discussion regarding
the theoretical calculation results of the electric dipole moment. One
of the authors (N. K.) thanks Mr. K. Ebihara from Mitsubishi Electric
Corp. for valuable comments on the device simulation procedure.

\section*{Author contributions}
Y.Y. and A.K. conceived and designed the experiments. A.K, N.K. and Y.W.
prepared SiC pn diodes, confirmed their electrical properties and
carried out device simulations. Y.Y. performed PBW for the
V\textsubscript{Si} array formation. Y.Y., A.K. and N.M. set up the
home-built confocal microscope. Y.Y. conducted electrical field sensing
measurements. R.A., S.D. and Y.N. carried out theoretical study. All
authors analyzed the data and discussed the results. Y.Y., A.K., R.A.
and S.D. wrote the manuscript with inputs from all the authors.

\newpage

\setcounter{figure}{0}
\setcounter{table}{0}
\renewcommand{\thefigure}{S\arabic{figure}}
\renewcommand{\thetable}{S\arabic{table}}

\begin{center}
\textbf{\large Supplementary Information}
\end{center}

\section*{Supplementary Information 1: TCAD device simulations of electric field.}

We used commercial TCAD device simulator (Synopsys Sentaurus Version
X-2025.06). Specifically, the internal electric field in the
edge-termination region was computed in the \emph{SDevice} module using
the Bank-Rose nonlinear solver with the ILS method. Material parameters
were adopted from the references  \cite{a1,a3}, and a p-type doping profile
measured by Secondary Ion Mass Spectrometry was incorporated into the
models. Two device configurations were modeled: (1) a standalone device
in isolation, (2) a device adjacent to a conductive wire biased at 0 V.

Figures S1a,b show the heat map of electric filed
(\emph{E\textsubscript{z}} and \emph{E\textsubscript{x}}) in the model
(1). The values of each electric field in the main text were extracted
from the result of the model (1). In the model (2), we then investigated
the influence of the enameled wire and Fluorinert on electric field
distribution by introducing a metallic conductor on the top of the
device varying the distance (\emph{D}) from the device surface, as well
as the relative dielectric constant ($\epsilon$\textsubscript{r}) between the
metal and device surface.

Figures S2a-d shows the simulated electric field with different
\emph{D}. Due to the electric potential being fixed to zero by the wire,
the electric field bends towards the periphery of the device. This
effect was also observed in the electric field inside the device,
becoming more pronounced on the surface and the periphery of the device.
When the wire was separated by 50 $\mu$m or more from the surface, the
effect was virtually eliminated at least within the device.

Another simulation was performed with the fixed \emph{D} (= 4 $\mu$m) and
different dielectric constants between the wire and the device surface.
For $\epsilon$\textsubscript{r}, we assume three different mediums; air ($\epsilon$ = 1.0
$\epsilon$\textsubscript{0}), Fluorinert ($\epsilon$ = 1.89 $\epsilon$\textsubscript{0}), and resin
($\epsilon$ = 3.8 $\epsilon$\textsubscript{0}), where $\epsilon$\textsubscript{0} is the dielectric
constant in vacuum. The results are shown in Fig. S3a-d. The larger
$\epsilon$\textsubscript{r} is, the more the electric field bends toward the
periphery of the device. We used the result of \emph{D} = 4 $\mu$m and
$\epsilon$\textsubscript{r} = 1.89 in the Fig. 5b,c in the main text. Noting
that, the effect of the wire and the relative dielectric constant
becomes negligible in the vicinity of the electrode, approximately
\emph{x} \textless{} 50 $\mu$m, therefore not affecting the introduction of
dipole moments discussed in the main text.

\begin{figure}[htb]
\centering{\includegraphics[width=12cm]{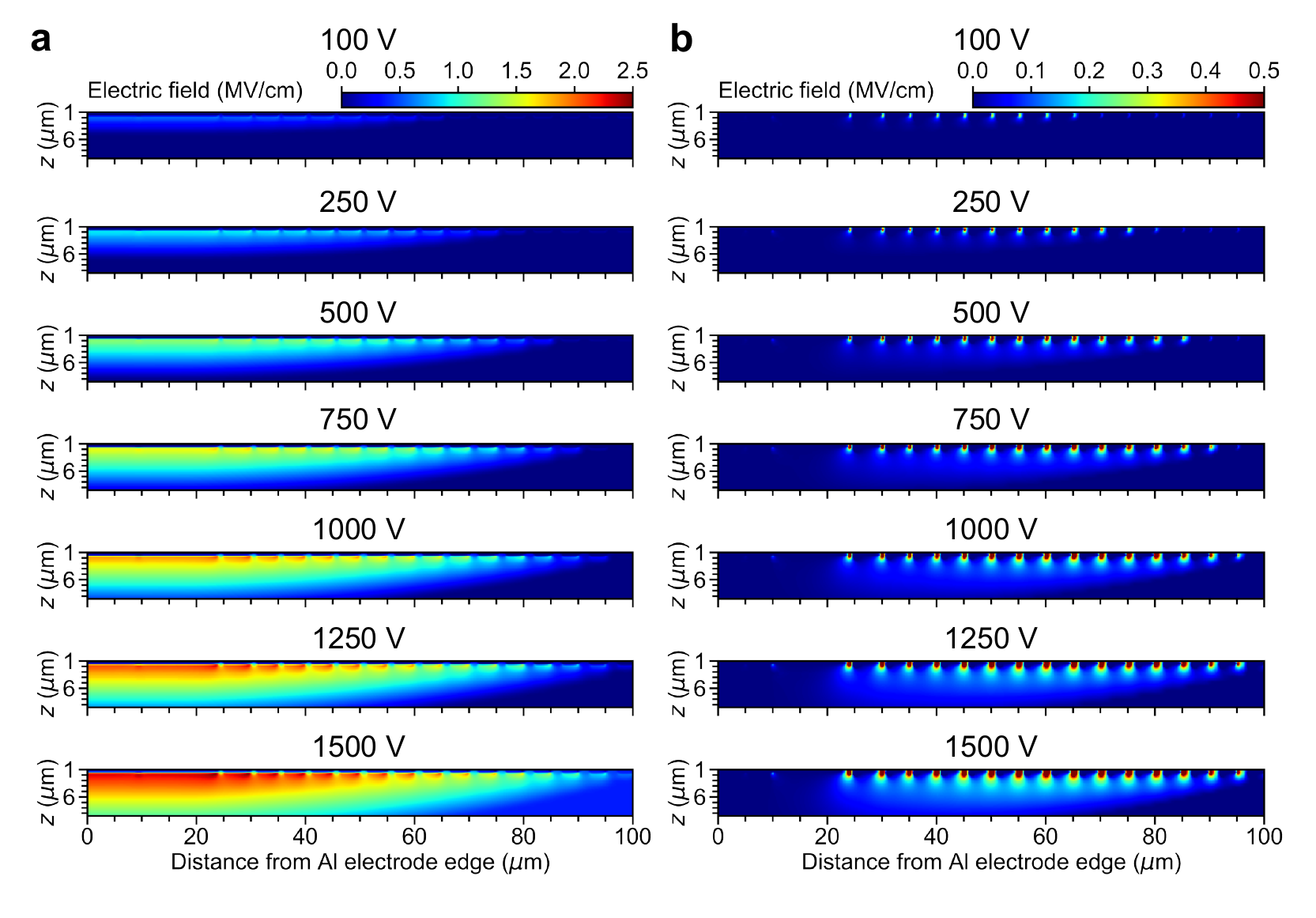}}
\caption{Simulated distribution of electric-field components.
\textbf{a}, \emph{E\textsubscript{z}} under reverse bias of 100-1500 V.
\textbf{b}, \emph{E\textsubscript{x}} under reverse bias of 100-1500 V.
In these simulations, the diode's edge termination region extends from
\emph{x} = 4 $\mu$m to 92 $\mu$m.}
\end{figure}

\begin{figure}[tb]
\centering{\includegraphics[width=12cm]{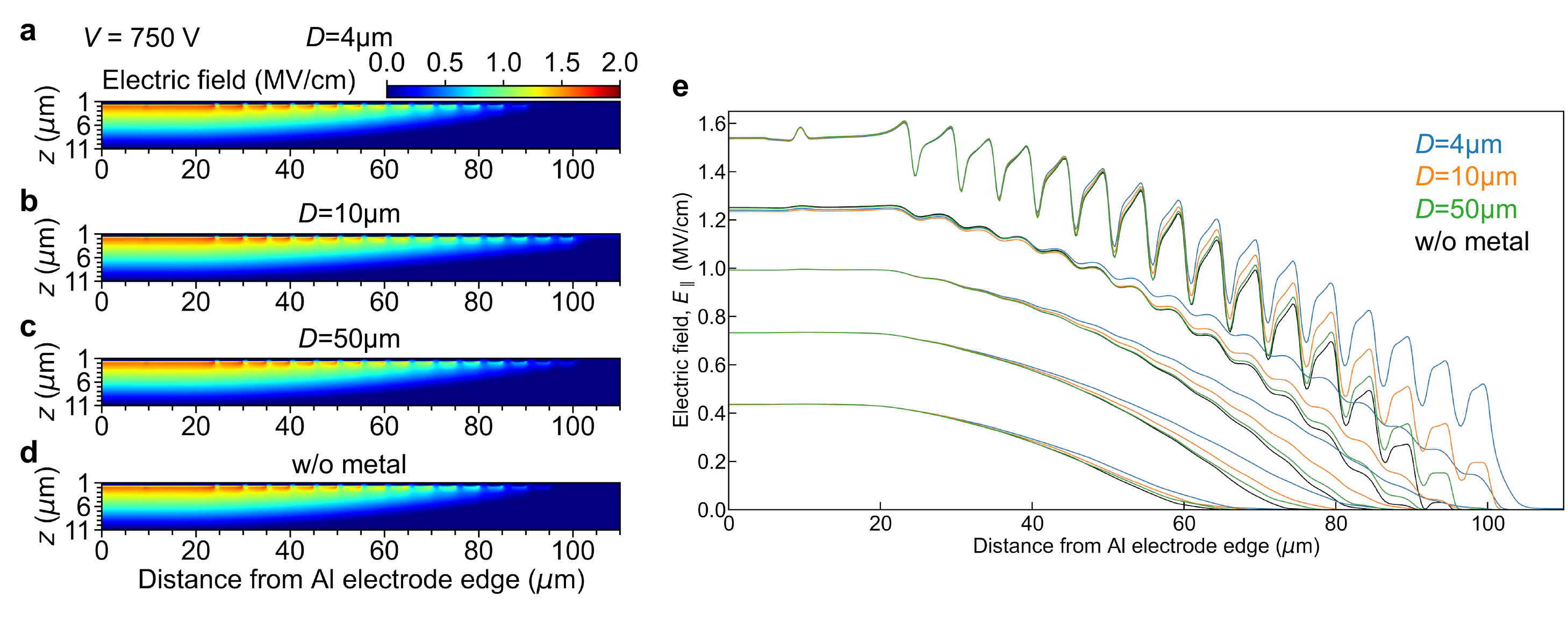}}
\caption{Simulation results of electric field
(\emph{E}\textsubscript{\textbar\textbar{}}) under reverse bias of 750 V
as a function of the distance (\emph{D}), while $\epsilon$\textsubscript{r} was
set to be 1.0. \textbf{a}, \emph{D} = 4 $\mu$m. \textbf{b}, \emph{D} = 10
$\mu$m. \textbf{c}, \emph{D} = 50 $\mu$m. \textbf{d}, Without metal. \textbf{e},
Extracted values at the specific depth of 2.1, 3.7, 5.0, 6.5, and 8.1
$\mu$m, respectively.}
\end{figure}

\begin{figure}[tb]
\centering{\includegraphics[width=12cm]{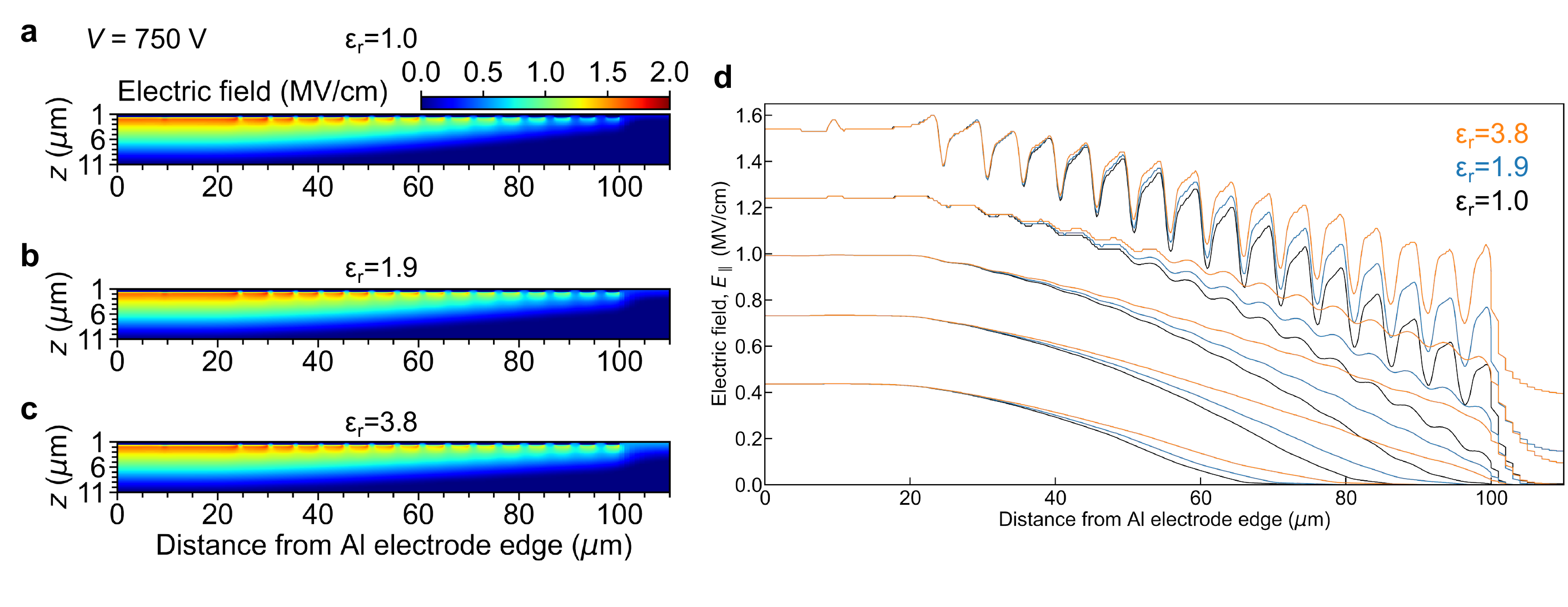}}
\caption{Simulation results of the electric field
(\emph{E}\textsubscript{\textbar\textbar{}}) under reverse bias of 750 V
at \emph{D} = 4 $\mu$m for different relative dielectric constants. \textbf{a}, $\epsilon$ =
1.0 $\epsilon$\textsubscript{0}. \textbf{b}, $\epsilon$ = 1.89 $\epsilon$\textsubscript{0}.
\textbf{c}, $\epsilon$ = 3.8 $\epsilon$\textsubscript{0}. \textbf{d}, Extracted field
values at depths of 2.1, 3.7, 5.0, 6.5, and 8.1 $\mu$m, respectively.}
\end{figure}

\section*{Supplementary Information 2: Sample setup.}

The pn diodes used in this study have a breakdown voltage exceeding 1500
V. Applying such high voltages in air causes surface discharge.
Therefore, as shown in Fig. S4, the sample and the printed circuit board
(PCB) including the DC and RF lines were immersed in an insulating
liquid, Fluorinert, to prevent the discharge. Since the refractive index
(= \emph{n}) of Fluorinert is 1.29, a water immersion objective lens was
selected to improve photon collection efficiency (cf. \emph{n} = 1.33
for water, \emph{n} = 1.52 for oil).

Figure S5 shows optical transmittance measurement results of the
Fluorinert used in this study. We confirmed approximately 95\%
transmittance in the 700-1100 nm range. There is almost no intensity
attenuation for the excitation laser (785 nm) and photoluminescence (PL)
detected by avalanche photodiode (APD) (900-1000 nm, with long-pass
filter (900nm) and APD detection efficiency). The fluctuations of data
in the 900-1100 nm range are attributed to the measurement apparatus.
Although PL characteristics of Fluorinert under 785 nm excitation laser
exposure were not investigated, the influence of PL from Fluorinert on
experiments is negligible because PL from V\textsubscript{Si} dot array
was clearly confirmed in the PL mapping image shown in Fig. 2g in the
main text, and ODMR measurements were feasible.

\begin{figure}[tb]
\centering{\includegraphics[width=11cm]{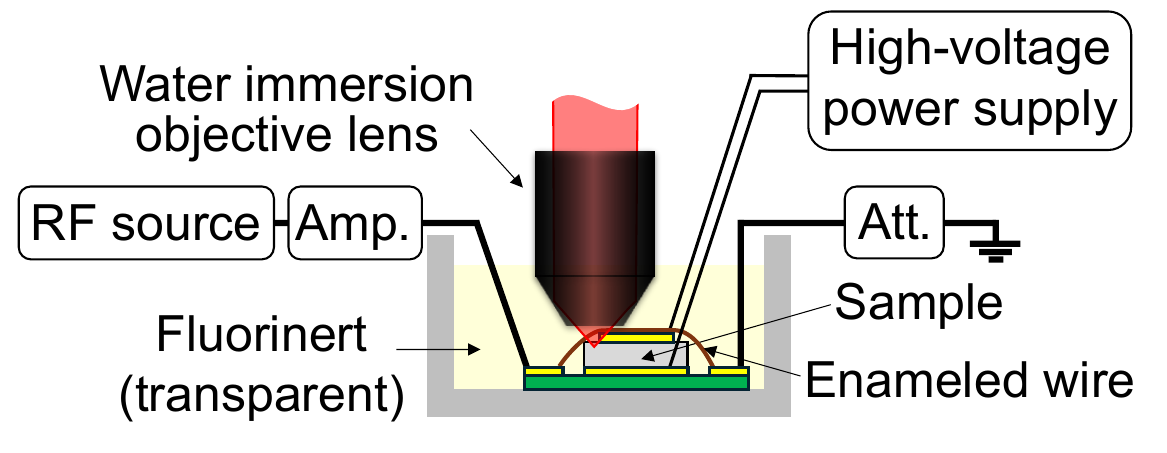}}
\caption{Schematic image of sample setup.}
\end{figure}

\begin{figure}[tb]
\centering{\includegraphics[width=8cm]{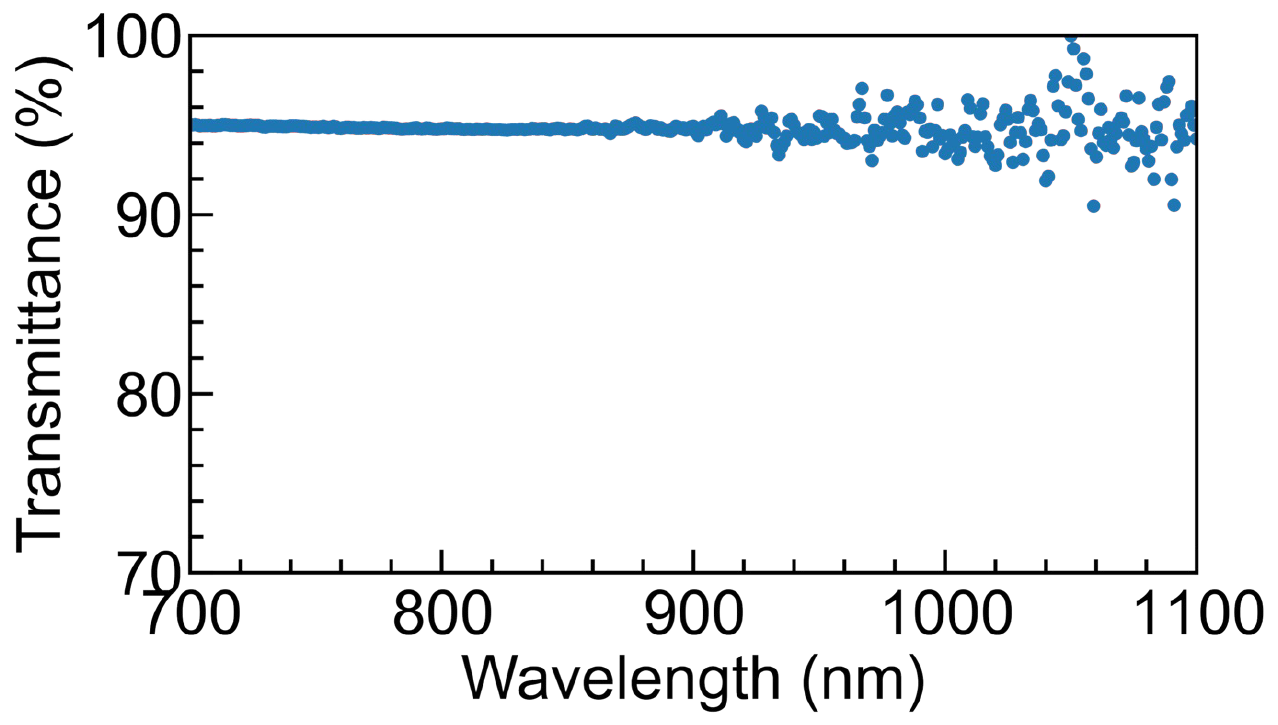}}
\caption{Optical transmittance of Florinate. The fluctuations of data
in the 900-1100 nm range are attributed to the measurement apparatus.}
\end{figure}

\section*{Supplementary Information 3: V\textsubscript{Si} density dependence of electric
field measurement.}

The sensitivity of quantum sensors is proportional to \(\sqrt{NT_{2}}\),
where N is the number of spin defects and T\textsubscript{2} is the spin
relaxation time  \cite{b3}. Therefore, increasing N without a decrease in
T\textsubscript{2} caused by spin-spin interactions is crucial for
maximizing sensor sensitivity. Hence, we investigated the
V\textsubscript{Si} density dependence of electric field measurements.
The results are shown in Fig. S6.
\emph{E}\textsubscript{\textbar\textbar{}} is the simulated values.
V\textsubscript{Si} density was estimated based on the reported and
experimentally determined V\textsubscript{Si} generation rate for H ion,
0.1 V\textsubscript{Si}/H \cite{a29}, and an enhancement factor of
V\textsubscript{Si} generation rate by changing from H to He ion
calculated by SRIM simulation (0.8 V\textsubscript{Si}/He).

As in Fig. 3b in the main text, a linear shift in the resonant frequency
was observed up to V\textsubscript{Si} density of $\sim$9$\times$10\textsuperscript{15} cm\textsuperscript{-3}. On the other hand, a
nonlinear change in the resonant frequency was observed for
V\textsubscript{Si} density more than $\sim$3$\times$10\textsuperscript{16} cm\textsuperscript{-3}. This result clearly shows
that there is an upper limit to the V\textsubscript{Si} density for
accurate electric field measurements. This limitation is not present in
magnetic field and temperature measurements and must be taken into
consideration in practical use. As shown in Fig. S7, device
characteristics remain unaffected even at 3$\times$10\textsuperscript{17}
cm\textsuperscript{-3} (= 3$\times$10\textsuperscript{5} He/$\phi$1$\mu$m) in the case of the particle
beam writing (PBW) using a dot pattern, suggesting that the nonlinear
change in the resonant frequency shown in Fig. S6 is not attributable to
device degradation.

\begin{figure}[htb]
\centering{\includegraphics[width=8cm]{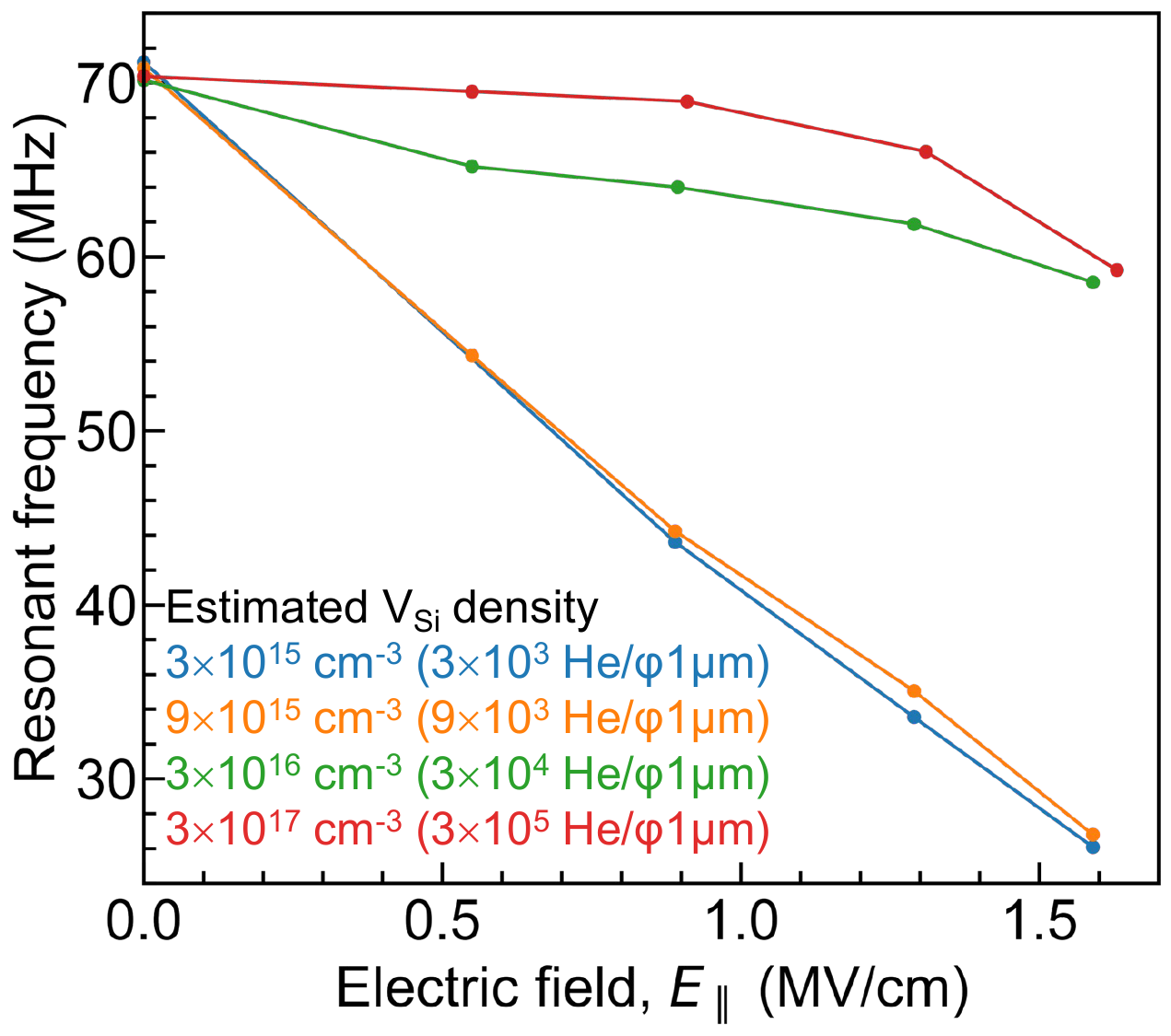}}
\caption{Electric field dependence of the resonance frequency as a
function of V\textsubscript{Si} density. Electric fields are derived
from device simulations. A transition from linear to nonlinear behaviour
in the resonance frequency was observed at V\textsubscript{Si} densities
more than 3$\times$10\textsuperscript{16} cm\textsuperscript{-3}.}
\end{figure}

\section*{Supplementary Information 4: Current-Voltage (I-V) characteristics.}

The influence of the V\textsubscript{Si} formation process (PBW) on
device characteristics was verified using two types of PBW patterns as
shown in Fig. S7a,d. Figures S7b-f show I-V characteristics of the pn
diodes used in this study. The ion energy for V\textsubscript{Si}
formation was set to 0.75 MeV. No remarkable change was observed under
forward bias (Fig. S7b,e) for all irradiation conditions. In the case of
reverse bias (Fig. S7c,f), the untreated sample exhibited breakdown at
approximately 1700 V, which is as designed. For a $\phi$1 $\mu$m dot pattern, no
significant differences were observed up to an ion fluence of 3$\times$10\textsuperscript{5} He/$\phi$1$\mu$m compared to the untreated sample,
confirming no degradation in device performance. In contrast, the
increase of reverse leakage current was confirmed for PBW with a fill
pattern even for an ion fluence of 3$\times$10\textsuperscript{3} He/$\phi$1$\mu$m,
which is attributed to the generation of various defects, including
V\textsubscript{Si}, introduced at the p/n interface in the edge
termination region by the PBW. These results clearly demonstrate that
selective formation of V\textsubscript{Si} is essential for evaluating
ideal device states using V\textsubscript{Si}-based quantum sensors.

\begin{figure}[tb]
\centering{\includegraphics[width=12cm]{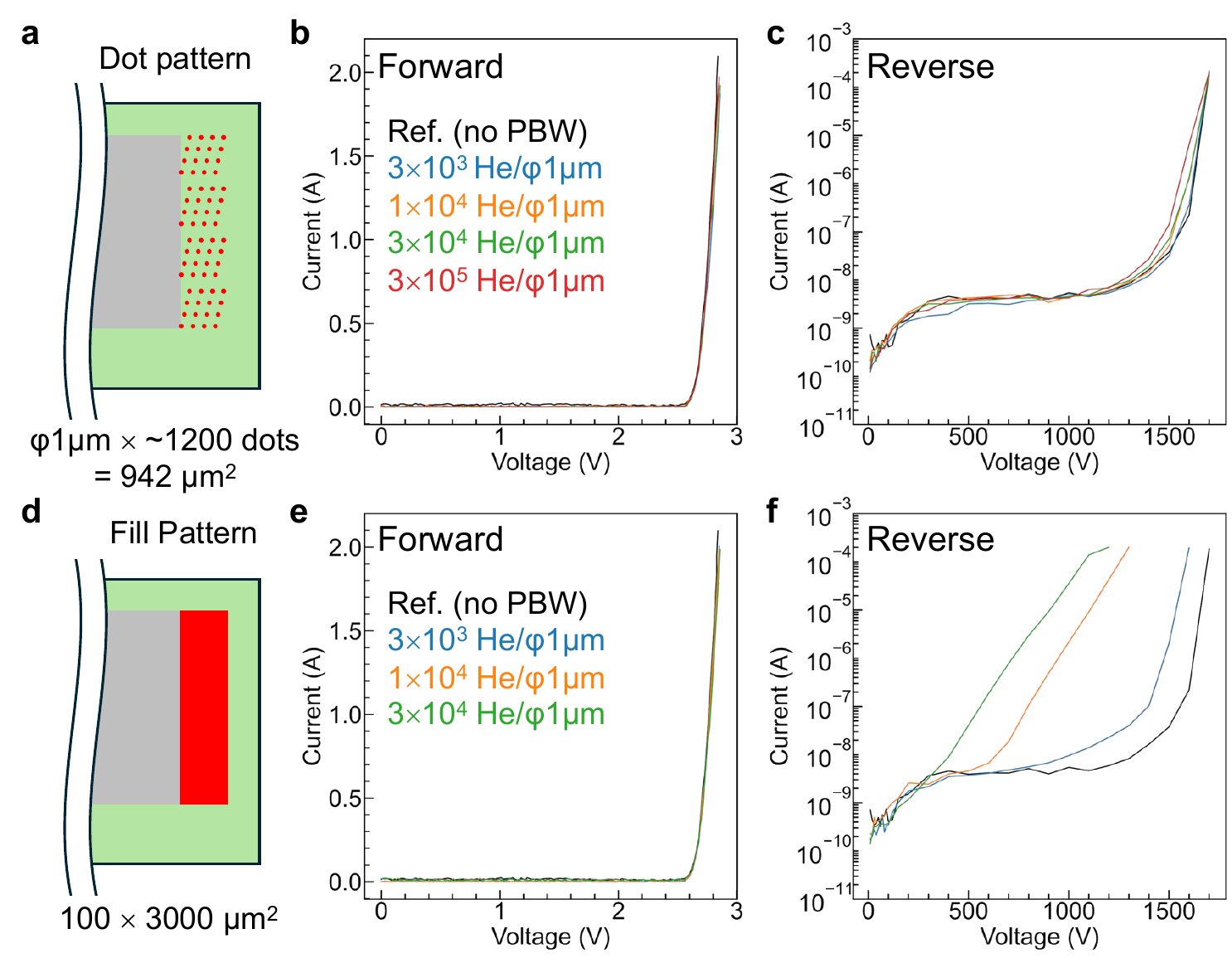}}
\caption{PBW patterns and I-V curves. \textbf{a-c}, Schematic of dot
pattern and corresponding I-V curves \textbf{d-f}, Schematic of fill
pattern and corresponding I-V curves.}
\end{figure}

\section*{Supplementary Information 5: Theoretical calculation of the electric dipole moments.}

We calculated \emph{d}\textsubscript{\textbar\textbar{}}/\emph{h} and
\emph{d}\textsubscript{$\bot$}/\emph{h} using the first-principles
calculation. The first-principles calculations were done with the VASP
package using the projector-augmented wave method for atomic
pseudopotentials \cite{a43, a44, a45, a46, a47}. The GGA-PBE functional \cite{a48} was taken
for approximating the exchange-correlation energy. Energy cutoff for the
plane waves was set to 500 eV. Only Gamma point was considered for the
self-consistent charge density and D-tensor calculations. The latter was
calculated using the Rayson-Briddon formula \cite{a49} as implemented in
VASP. Analyses of the planar-averaged electrostatic potential were
performed with VASPKIT \cite{a50}.

We made a supercell slab of 4H-SiC in a vacuum, which include 5$\times$5$\times$2
units of (SiC)\textsubscript{4} (Fig. S8a-c). To remove the artificial
in-gap states due to the surface dangling bonds, we terminated the
surfaces by radicals. The -OH and -H radicals were attached to the Si
and C surfaces, respectively, which were helpful to remove artificial
in-gap surface states due to the difference in the polarizability of Si
and C.

First, we prepared the slab using the experimental crystal structure
\cite{b13} and optimized it allowing only the surface radicals to move.
The -H radicals remained on top of the surface C atoms, whereas the -OH
radicals uniformly bent toward the surface, by which the
\emph{C\textsubscript{3v}} symmetry of the system was broken. See Fig.
S8a-c for the optimized structure. We next calculated the intrinsic
electric fields in the middle of the slab, which corresponded to those
from the macroscopic estimate in Main. We display in Fig. S9a the
calculated planar-averaged Hartree plus ionic potential, which include
the externally applied scalar potential, and its average in \emph{z}
direction, whose slope gives us the electric field. Inside the slab, the
potential shows nonzero slope, which indicates nonzero electric field as
a consequence of the different polarizabilities of Si and C. Contrary to
the bulk result as shown in Fig. S6b, the electronic band structure with
the (-OH, -OH) termination (Fig. S9d) was found to host in-gap bands
above the valence top, which appears highly dispersive even in the
supercell Brillouin zone; they are artificial surface states. In the
case of the (-H, -OH) termination, the local potential minimum at the
top surface (\emph{z} $\sim$ 50\AA) was absent as seen in Fig.
S9a. Thanks to this, the surface states were well removed from the gap
(Fig. S5c), as well as the overall band structure is similar to the bulk
one. We also performed the Hartree+ionic potential calculation with
external field \emph{EFIELD} = 0.0, -0.025 and -0.05 (eV/\AA) as
implemented in VASP and calculated the total electric fields from the
``Averaged'' slopes of the potential as shown in Fig. S9a (see Table S1
for the values). These three values were used for calculating the linear
response coefficient of the \emph{D}-tensor below.

\begin{figure}[tb]
\centering{\includegraphics[width=11cm]{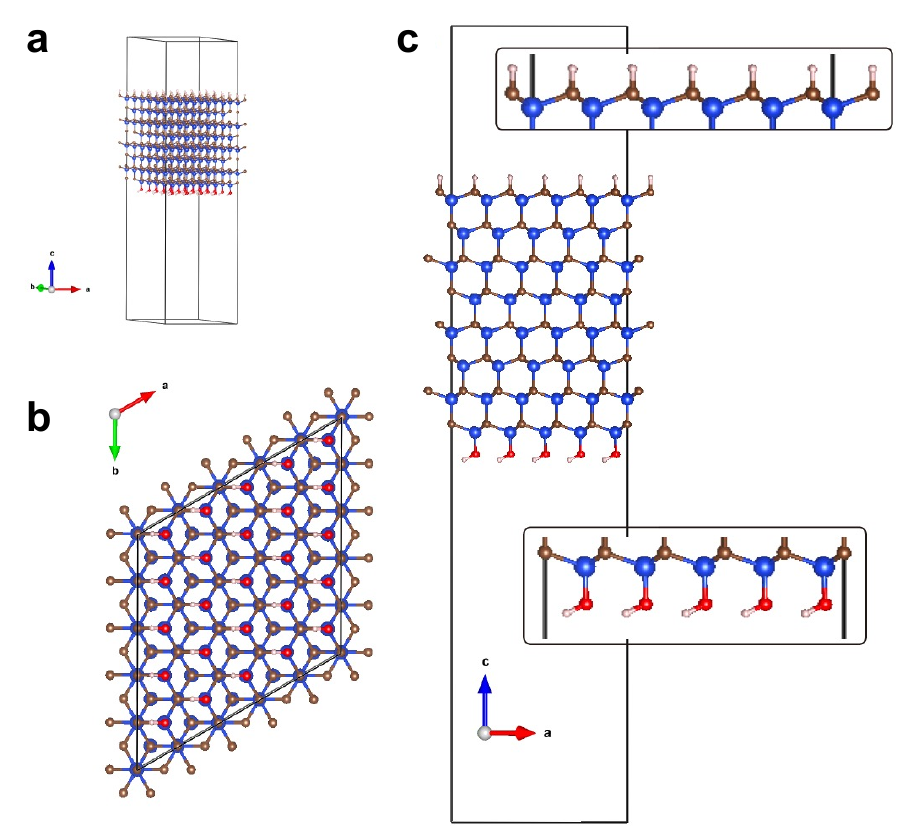}}
\caption{Schematic figure of the slab structure used for the
first-principles calculations. \textbf{a}, Structure before introducing
the defects. \textbf{b,c}, Its bottom and side views. The insets show
the closeup view of the top and bottom layer terminated by -H and -OH
radicals, where the optimized bond lengths were 1.104\AA, 1.670\AA and
0.977\AA  for the H-C, Si-O and O-H bonds, and bond angle for Si-O-H was
114.1 degree.}
\end{figure}

From the (-H, -OH) slab thus optimized, we removed one \emph{k}-site Si
from the slab to make the vacancy V\textsubscript{Si} and replaced one
Si with P as an artificial donor. The latter P\textsubscript{Si} was
introduced in a distant location with the same \emph{z}-position (Fig.
S10a-c), by which we aimed to minimize the effect of artificial field
exerted on V\textsubscript{Si} by the positively charged
P\textsubscript{Si} and its translated images in the planar directions.
Varying the external field, we again optimized the positions of atoms
around V\textsubscript{Si} up to the second nearest neighbors and
examined the band structures allowing spin polarization (Fig. S10d,e).
The in-gap defect states indeed emerged, which form the S=3/2 quartet.
With the near-vacancy relaxation (Fig. S10e), we observed small downward
shifts of the defect states and slightly dispersive bands at the valence
top; they might affect the \emph{D}-tensor. We calculated the
field-dependent \emph{D}-tensor with and without the near-vacancy
optimization. The results are summarized in Table S1. The ZFS values
calculated from \emph{D\textsubscript{zz}} were generally larger than
the experimentally measured ones (10-70 MHz), which might be due to the
artificial residual \emph{E}\textsubscript{eff} in polar semiconductors
that would be relaxed in the experimental samples by structural
reconstruction and the attached substrate. Nevertheless, the
linear-response coefficient \emph{d}\textsubscript{\textbar\textbar{}},
which should be robust against the zero of \emph{E}\textsubscript{eff},
agrees well with the experiment.

We further estimated the response to the in-plane electric field,
\emph{d}\textsubscript{$\bot$}/\emph{h}, exploiting the present artificial
slab setup. Since the slab contains
V\textsubscript{Si}\textsuperscript{-},
P\textsubscript{Si}\textsuperscript{+} and their periodic images, the
target V\textsubscript{Si}\textsuperscript{-} feels an electric field
exerted from the other charged defects, which may be utilized as sources
of the field parallel to the slab plane. However, in the Kohn-Sham
calculation the total scalar potential includes the field from those
surroundings, as well as that induced by the target
V\textsubscript{Si}\textsuperscript{-}, which cannot be separated. We
estimated the former by adopting a very simple model that the
surrounding defects are all point charges placed within uniform medium
with dielectric constant 10.0 ($\approx$experimental value for bulk 4H-SiC), by
which we can compare the D-tensor to the in-plane electric field. Also,
we calculated the D-tensor with a slight shift to P\textsubscript{Si}
for linear response. Comparing the changes of the in-plane fields and
(\emph{D\textsubscript{xx}}, \emph{D\textsubscript{yy}},
\emph{D\textsubscript{xy}}), we estimated \emph{d}\textsubscript{$\bot$} $\approx$
+16 (MHz/(MeV/cm)).

\begin{figure}[H]
\centering{\includegraphics[width=11cm]{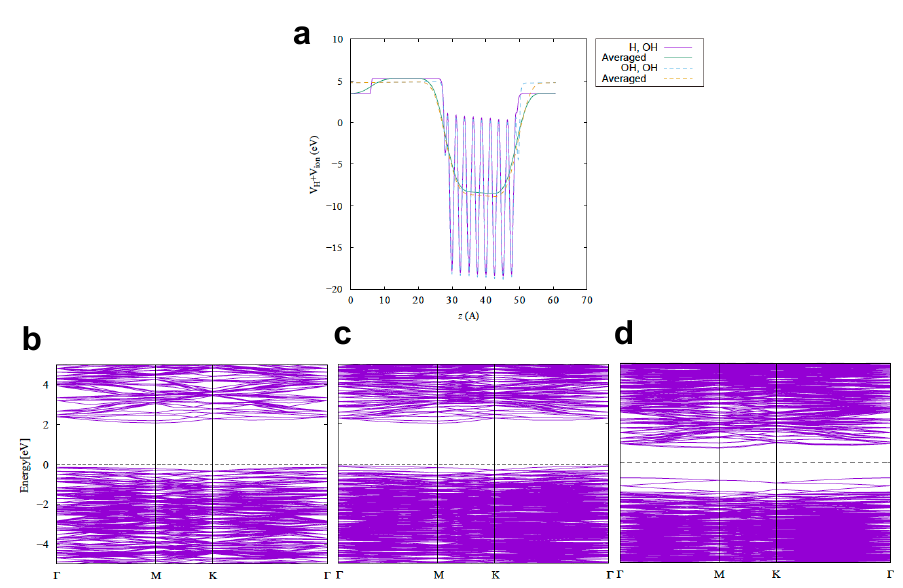}}
\caption{Effect of surface termination on the electronic structure in
the pristine slab. \textbf{a}, Planar-averaged Hartree plus ionic
potential for the different terminations. The ``Averaged'' data are
calculated by averaging the values within the width 5.08\AA using VASPKIT
\cite{a48}. For the (-OH, -OH) termination the dipole correction \cite{b14, b15}, represented by the step in the vacuum region, was applied.
\textbf{b-d}, Band structure of the bulk 4H-SiC folded to the hexagonal
Brillouin zone of the (5 5 2) supercell, slab with termination (-H, -OH)
and (-OH, -OH).}
\end{figure}

\begin{figure}[H]
\centering{\includegraphics[width=11cm]{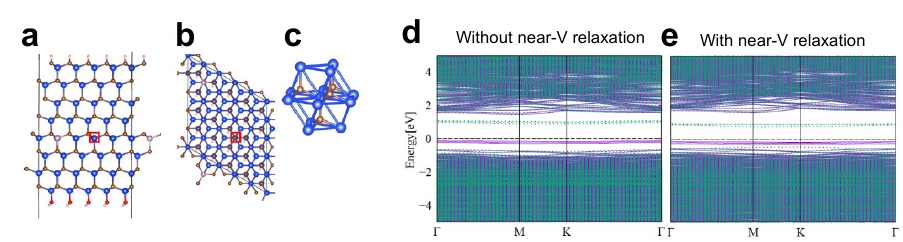}}
\caption{Calculation slab with \emph{k}-V\textsubscript{Si} (red
box) and P\textsubscript{Si} (pale purple) and its band structures.
\textbf{a}, Side view of the slab. \textbf{b}, Top view of the layer
where both defects reside. \textbf{c}, Neighboring geometry of the
k-site up to the second nearest. \textbf{d}, Calculated spin-polarized
band structure of the slab with \emph{EFIELD} = 0.0, where up-(down-)
spin bands are drawn in purple (green). \textbf{e}, The band structure
with optimization of the atomic positions up to the second nearest to
\emph{k}-V\textsubscript{Si}.}
\end{figure}

\begin{table}[H]
 \caption{The input \emph{EFIELD} parameter used for VASP and calculated
effective field \emph{E}\textsubscript{eff}, \emph{D}-tensor component
\emph{D\textsubscript{zz}}, zero-field splitting (=
3\emph{D\textsubscript{zz}}) and
\emph{d}\textsubscript{\textbar\textbar{}}. The sign of
\emph{E}\textsubscript{eff} is defined consistent with the main text.
Note that \emph{D\textsubscript{zz}} here is defined with the spin
Hamiltonian \\
\(\mathcal{\bm{H}}_{\bm{ss}}\mathbf{=}\left( \bm{S}_{\bm{x}}\mathbf{,\ \ }\bm{S}_{\bm{y}}\mathbf{,\ \ }\bm{S}_{\bm{z}} \right)\mathbf{diag}\left\{ \mathbf{-}\frac{\bm{D}_{\bm{zz}}}{\mathbf{2\ }}, \mathbf{-}\frac{\bm{D}_{\bm{zz}}}{\mathbf{2\ }}, \bm{D}_{\bm{zz}} \right\}_{}^{\bm{t}}\left( \bm{S}_{\bm{x}}\mathbf{,\ \ }\bm{S}_{\bm{y}}\mathbf{,\ \ }\bm{S}_{\bm{z}} \right)\mathbf{=}\frac{\mathbf{3}}{\mathbf{2}\bm{D}_{\bm{zz}}\bm{S}_{\bm{z}}^{\mathbf{2}}}\mathbf{+ const.}\):\\
It differs from that defined in Ref.23 in the main text by factor 2/3. The values of \emph{d}\textsubscript{\textbar\textbar{}} in the last row includes the correction ($\times$3/2) for this for better comparisons.
}
 \centering
  \begin{tabular}{|c|c|c|c|c|c|c|}
  \hline
     & \multicolumn{3}{|c|}{Without near-V relaxation} & \multicolumn{3}{|c|}{With near-V relaxation} \\
   \hline
    \emph{EFIELD} (eV/\AA) & 0.0 & -0.025 & -0.05 & 0.0 & -0.025 & -0.05 \\
   \hline
   \emph{E}\textsubscript{eff} (MeV/cm) & 3.438 & 3.096 & 2.755 & 3.438 & 3.096 & 2.755 \\
   \hline
   \emph{D\textsubscript{zz}} (MHz) & 42.567 & 45.774 & 49.008 & 26.034 & 29.582 & 33.142 \\
   \hline
   ZFS (MHz) & 127.701 & 137.322 & 147.024 & 78.102 & 88.746 & 99.426 \\
   \hline
   \emph{d}\textsubscript{\textbar\textbar{}} (MHz/(MeV/cm)) & \multicolumn{3}{|c|}{-14.1} & \multicolumn{3}{|c|}{-15.6} \\
   \hline
  \end{tabular}
\end{table}

\section*{Supplementary Information 6: Determination of V\textsubscript{Si} spot position.}

To determine the electric field around the V\textsubscript{Si} used in
the measurements based on device simulation results, the determination
of each V\textsubscript{Si} spot position is crucial. For the \emph{x}
coordinate, the distance between the edge of the Al electrode and
V\textsubscript{Si} spot closest to the electrode was determined using a
PL mapping image. The high positional accuracy of our PBW system, less
than 100 nm, ensures that the distance between each V\textsubscript{Si}
spot is exactly 20 $\mu$m. For the \emph{z} coordinate, the resolution of
our confocal microscope is not high enough to determine it accurately
from a PL mapping image (\emph{xz} scan). In this study, we decided to
use the V\textsubscript{Si} depth profile obtained from stopping and
range of ions in matter (SRIM) simulation. To examine its validity, the
reference \emph{z} coordinate was experimentally determined using both
optically detected magnetic resonance (ODMR) and cathodoluminescence
(CL) measurements near the p/n interface.

Near the p/n interface, the electric field applied to the p-type
epilayer is several orders of magnitude smaller than that for the n-type
epilayer. In this situation, no electric field response from
V\textsubscript{Si}-based quantum sensor formed within the p-type
epilayer are expected to be observed. ODMR spectrum consisting of both
zero-response signal and specific electric field response signal should
be obtained from the V\textsubscript{Si} formed across the p/n
interface. Therefore, we performed ODMR measurements using
V\textsubscript{Si} spots formed at different depths near the p/n
interface.

Figure S11 shows depth profile of V\textsubscript{Si} calculated by SRIM
simulations. The simulation shows that V\textsubscript{Si} ensemble with
a full width at half maximum of 0.2-0.4 $\mu$m in the depth direction is
formed by He ions. At an ion energy of 0.4 MeV, V\textsubscript{Si}
ensemble distributed only in the p-type epilayer is expected to form,
and at 0.5 MeV, the distribution of V\textsubscript{Si} ensemble across
the p/n interface is predicted. At ion energies exceeding 0.75 MeV, ion
energies primarily used in this study, V\textsubscript{Si} ensemble is
located almost entirely in the n-type epilayer. Figures S12a-c show ODMR
spectra obtained from V\textsubscript{Si} ensemble at three different
depths. A single peak with resonance frequency of $\sim$70 MHz
is observed even at reverse bias of 750 V for 0.4 MeV (Fig. S12a),
indicating the absence of an applied electric field as expected in the
p-type epilayer. For 0.5 MeV, we observed a mixture of two ODMR signals
with resonant frequencies of $\sim$40MHz and $\sim$70
MHz (Fig. S12b). The latter is the ODMR signal observed in a zero
electric field. This result clearly indicates that V\textsubscript{Si}
ensemble was formed on both the p-type and n-type epilayers as predicted
by the SRIM simulation. At 0.75 MeV, the single peak with resonant
frequency of $\sim$40 MHz was observed at reverse bias of 750 V
(Fig. S12c), indicating the presence of V\textsubscript{Si} ensemble in
the n-type epilayer.

We performed CL measurements to experimentally confirm
V\textsubscript{Si} depth at an ion energy of 0.5 MeV. To increase CL
signal intensity, the ion fluence was set to 3$\times$10\textsuperscript{6}
He/$\phi$1$\mu$m, which is more than 100 times higher than the values used in the
electric field measurements. The D-center, which has been identified as
a defect-derived signal in photoluminescence measurements \cite{b16}, was
used as an indicator of the V\textsubscript{Si} depth. Figure S13 shows
the integral intensity of CL peak from the D-center as a function of
distance from the sample surface. The integral intensity reached its
maximum at \emph{z} $\sim$ 1.6 $\mu$m. This is consistent with
the SRIM and ODMR results mentioned above.

Based on these results, we concluded that the V\textsubscript{Si} depth
predicted by the SRIM simulation agrees well with the experimentally
verified values. Therefore, in this study, the peak positions of depth
profile calculated by the SRIM simulation were used as the \emph{z}
coordinate of V\textsubscript{Si} spot.

\begin{figure}[tb]
\centering{\includegraphics[width=7cm]{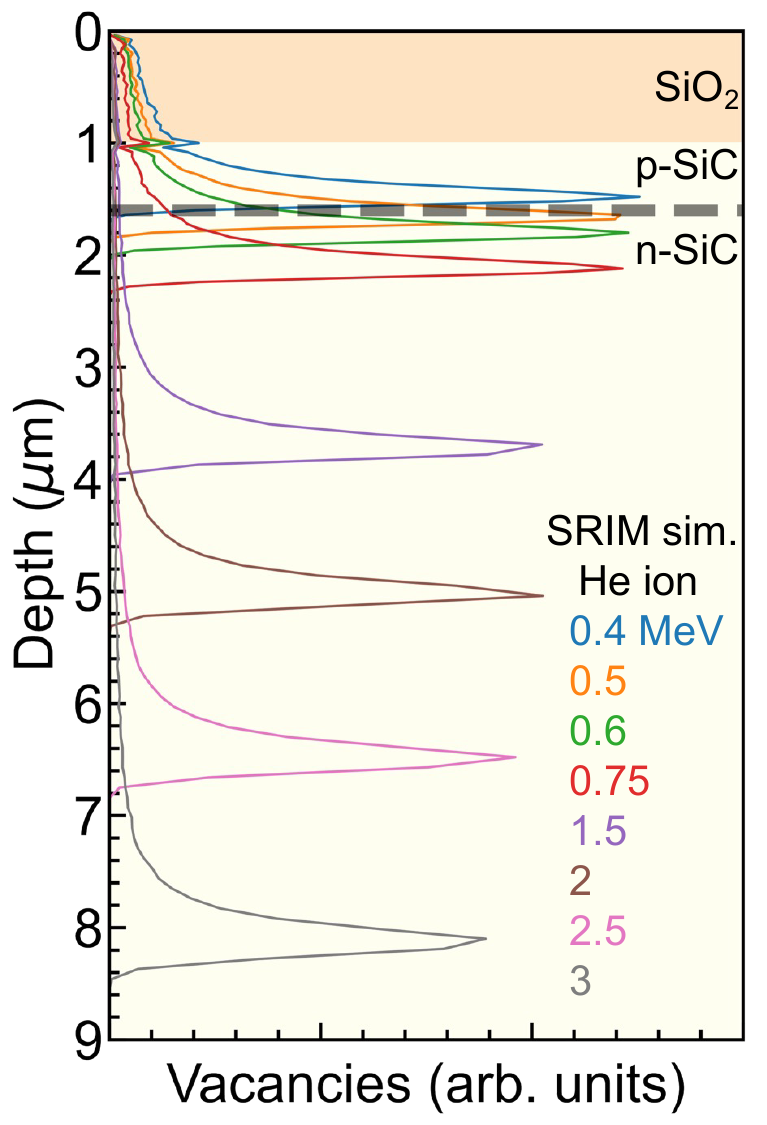}}
\caption{Depth profile of V\textsubscript{Si} calculated by SRIM
simulations. The dashed line indicates the p/n interface estimated from
the results of ODMR (Fig. S12a,b) and CL (Fig. S13).}
\end{figure}

\begin{figure}[tb]
\centering{\includegraphics[width=11cm]{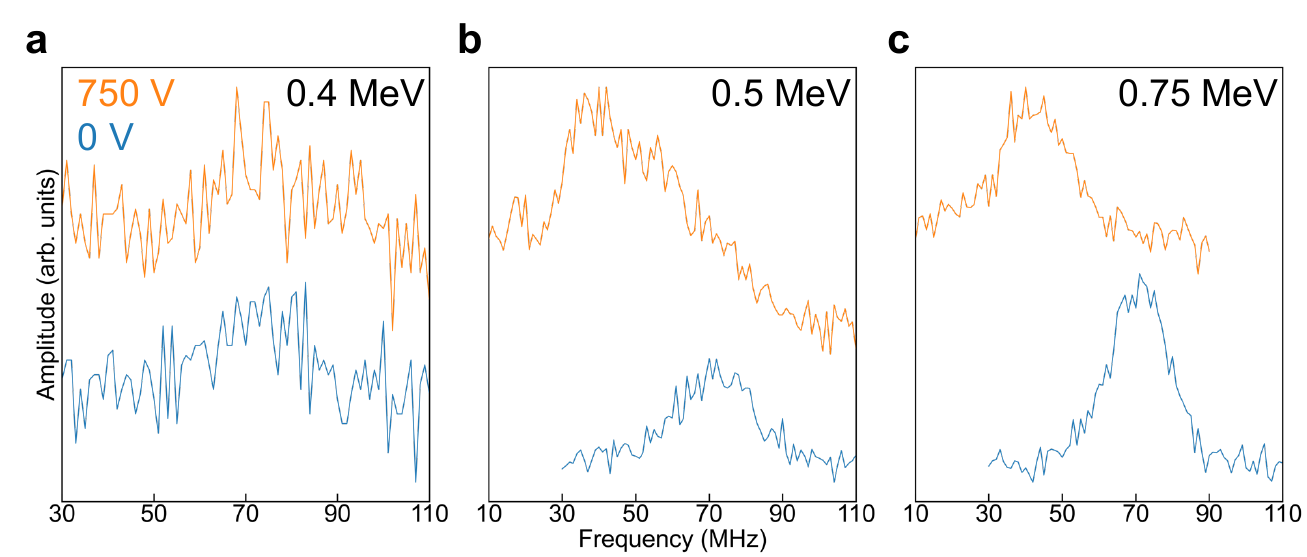}}
\caption{ODMR spectra obtained from V\textsubscript{Si} ensemble at
three different depths. \textbf{a}, For 0.4 MeV. \textbf{b}, For 0.5 MeV.
\textbf{c}, For 0.75 MeV.}
\end{figure}

\begin{figure}[H]
\centering{\includegraphics[width=6cm]{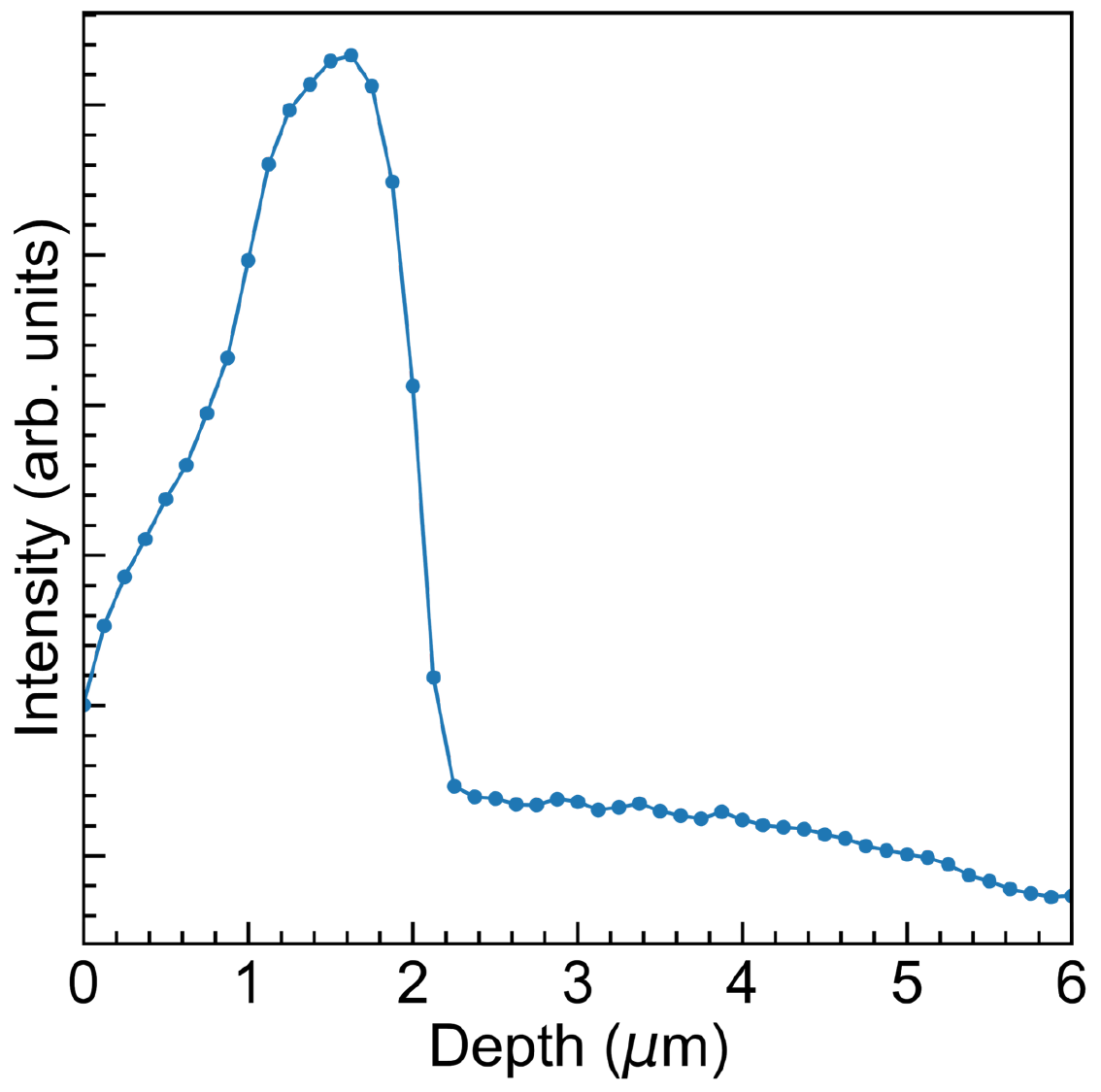}}
\caption{Integral intensity of CL peak from the D-center as a
function of distance from the sample surface.}
\end{figure}

\bibliographystyle{apsrev4-2}
\bibliography{bib1}

\end{document}